\documentclass{aa}
\usepackage{natbib}
\usepackage{graphics}
\usepackage[mathcal]{eucal}
\usepackage{amssymb} 

\newcommand{\ds}{\displaystyle}
\newcommand{\inv} {\frac {1}}

\long\def\jumpover#1{{}}

\newcommand{\deriv} [2] {\frac {d #1 } {d #2} }

\newcommand{\fig}[3]{
      \begin{figure}
        \resizebox{\hsize}{!}{\includegraphics  {#1}}
        \caption{#2}
        \label{#3}
        \end{figure} }
\newcommand{\eqn} [1] {
\begin{equation}#1
\end{equation}}
\newcommand{\eqna} [1] {
\begin{eqnarray}#1
\end{eqnarray}}

\begin{document}

\title{Numerical constraints on the model of stochastic excitation\\
of solar-type oscillations. }

\author{Samadi R. \inst{1,2} \and  
Nordlund {\AA}. \inst{3} \and Stein R.F. \inst{4} 
\and Goupil M.J. \inst{2} \and  Roxburgh I.  \inst{1,2} }

\institute{
Astronomy Unit, Queen Mary, University of London, London E14NS, UK \and
Observatoire de Paris, LESIA, CNRS UMR 8109, 92195 Meudon, France \and 
Niels Bohr Institute for Astronomy Physics and Geophysics, Copenhagen , Denmark\and
Department of Physics and Astronomy, Michigan State University, East Lansing, Michigan, USA
}
\offprints{R. Samadi}
\mail{Reza.Samadi@obspm.fr}
\date{\today} 

\abstract{ Analyses of a 3D simulation of the upper layers of
a solar convective  envelope provide 
constraints on the physical quantities
 which enter the  theoretical formulation of a stochastic excitation model 
of  solar p~modes,  for instance the convective velocities
and  the turbulent kinetic energy spectrum. 
These constraints are then used to compute   the acoustic excitation rate
 for solar p~modes, $P$.
The resulting values are  found   $\sim 5$ times larger 
than the values resulting from a computation in which 
convective velocities and entropy fluctuations are  obtained with a  1D solar envelope model 
built with the time-dependent, nonlocal  \citet{Gough77} extension
 of the mixing length formulation for convection (GMLT). \\
This difference is mainly due to the assumed mean 
anisotropy properties of the velocity field in the excitation region. The 3D
simulation suggests much larger horizontal  velocities compared to vertical
ones than in the 1D GMLT solar model.
The  values of $P$ obtained with the 3D simulation constraints
however are still too  small compared 
with the values inferred from solar observations.\\
Improvements in the description of the turbulent kinetic energy
spectrum and its depth dependence
 yield further increased  theoretical values of $P$ which bring them closer 
to the observations. It is also found that the source of excitation  
arising from  the advection of the turbulent fluctuations of entropy 
by the turbulent movements contributes $\sim 65-75 \%$   to the excitation
and therefore  remains
 dominant over the Reynolds stress contribution.
The  derived theoretical values of $P$ 
obtained with the 3D simulation constraints
 remain smaller by a factor  $\sim 3$ 
 compared with the solar observations. This shows that 
the stochastic excitation model still needs to be improved.
}

\maketitle

\section{Introduction}

Solar-type oscillations are believed to be stochastically excited by turbulent 
convection in the near-surface layers of the star. 
The excitation is caused by turbulent convective motions 
 which generate  acoustic energy which in turn is injected into the p~modes
\citep[e.g.][]{GK77}.
Measurements of the  acoustic energy injected into solar-like 
oscillations are among the goals of future space seismic missions such
 as the  COROT \citep{Baglin98}  and Eddington \citep{Favata00} missions.
These seismic data will  make it possible to constrain the theory of the
oscillation excitation and damping, to provide valuable information
about the properties of stellar convection, and hence
to severely  constrain stellar models.

Models for stochastic excitation of stellar p~modes have been  
proposed by several authors, \citep[e.g.][]{GK77,Osaki90,Balmforth92c,GMK94,Samadi00I}.
These semianalytical approaches 
yield the acoustic energy injected into solar-like 
oscillations. 
This offers the advantage of
testing separately several properties entering the excitation mechanism
which are not well understood or modeled.

Such  approaches require simplifying assumptions
which need to be validated before they can be used with confidence. 
They require an accurate knowledge of the properties of turbulent convection and,
unfortunately, current observations of the
solar granulation cannot provide a
determination of the turbulent spectrum precise enough in the present context
\citep{Rieutord00,Nordlund97}.
On the theoretical side,  theoretical models of turbulent convection, such as the mixing-length approaches or multiple size eddies approaches \citep[e.g.][]{Canuto91,Canuto96}, 
 provide a too limited description of
the characteristic scale length of the solar turbulent spectrum.

These theoretical formulations of stochastic excitation also involve 
scaling parameters which are determined to by the requirement that
the computed values of the oscillation 
amplitudes give the best fit to the solar seismic measurements \citep[e.g.][Paper~II hereafter]{Houdek99,Samadi00II}.
 When the scaling parameters are 
so adjusted, constraints and 
validation on the turbulent stellar medium
can only come from  seismic observations of other stars. 
Such accurate  data on the excitation rates 
 for other stars  than the Sun are not yet available.

 An alternative way  is then to consider 
results from 3D numerical simulations. They indeed 
enable one  to compute directly the rate at which p~modes 
are excited (e.g. this was undertaken for the Sun by \citet{Stein01II}).
Such methods are time consuming and do not easily allow 
massive  computations of the excitation rate for stars  
with different temperatures and luminosities. 
 They can provide quantities which  can be implemented
in  a  formulation for the excitation rate $P$. 
In any case we cannot avoid to use a 1D~model for computing accurate 
eigenfrequencies for the whole observed frequency range.

The purpose of the present paper is to provide 
a better insight  into the  excitation model
with a  semianalytical approach   but 
using a model of turbulence and values of
the  scaling parameters
derived from a 3D simulation of the solar outer layers.
We consider in this work the theoretical formulation of stochastic excitation
by \citet[hereafter Paper~I, see also \citet{Samadi01} for a detailed
summary]{Samadi00I}  which  includes a detailed treatment of turbulent convection.
This formulation involves two scaling parameters
which are  related to the spatial and temporal  
characteristics of the turbulence model.
Our final goal is  to test the excitation model without adjusting these parameters
and without the use of the mixing-length approach for estimating convective
velocities and entropy fluctuations.


\vskip 0.2 truecm

The paper is organized as follows: in Sect.\ref{sec:Theory of stochastic excitation}
we briefly recall
 the  adopted  formulation  for estimating the rate at which 
turbulent convection supplies energy to the p~modes (excitation rate $P(\nu)$). 
We emphasize  some  assumptions and approximations
entering this formulation.

In Section~\ref{sec:Constraints from the 3D simulation},
 a  3D numerical simulation of the upper part of the solar convection zone 
is used in order to determine the time averaged    
properties of turbulent convection: this provides   
constraints on the ingredients involved in the 
theoretical expression of the excitation rate, 
such as  scaling parameters,  velocity anisotropy factor,
the values of convective velocities and entropy fluctuations 
and  the $k$ (wavenumber) dependence of the kinetic turbulent spectrum.

These  constraints are then 
used in Section~\ref{sec:Consequences in term of p modes excitation}
 to compute the excitation rate $P(\nu)$,  for radial solar p~modes.
The results  are compared with  solar seismic observations as given
 in \citet{Chaplin98} and with a 1D mixing-length model built according
 to \citet{Gough77}'s non-local formulation of the mixing-length theory
 (GMLT hereafter).
In Section~\ref{sec:Conclusion} we summarize our results and
 discuss  some possible origins of the  remaining discrepancies 
with solar seismic observations and results by \citet{Stein01II}.

\section{Stochastic excitation}
\label{sec:Theory of stochastic excitation}

\subsection{The excitation model}

The rate at which  turbulent motions of the convective elements supply
energy to acoustic oscillation modes is computed as in 
Paper~I. For a given mode with eigenfrequency $\omega_0$, the excitation rate
 can be  written as (Eq.~58 and Eq.~59. of Paper~I) :
\eqna{
 P(\omega_0) = P_R + P_S
\label{eqn:P}
}
where 
\eqna{
P_{R,S} & = & \frac{ \pi^{3}  }
{ 2   I} \int_{0}^{M} {\rm d} m \, \rho_0  \, \frac{\Phi}{3}  \, w^4 \, F_{R,S} 
\label{eqn:P_RS}
}
where $\rho_0$ is  the mean averaged density, 
$w$ is the rms value of the vertical component of the velocity,
\eqn{
I \equiv   \int_0^{M} dm \,   \xi_r^2 
\label{eqn:I} 
} is the mode inertia, $\displaystyle{\xi_{\rm r}}$ is the
 radial component of the fluid displacement adiabatic eigenfunction $\vec\xi$,
 and $\Phi$ is an anisotropy factor.
Following \citet{Gough77}, we define
\eqn{
\Phi(z) \equiv 
\frac{\overline{ <{\vec u}^2> - <\vec{u}>^2}}{w^2} 
\label{eqn:phiz}
}
where $\vec u$ is the velocity field,  $< . >$ denotes horizontal  average  and $\overline{()}$  denotes time average.
The mean vertical velocity, $w$, is defined as:
\eqn{
w^2 \equiv \overline{ <u_z^2>-<u_z>^2 }
}
$P_{R}$,  $P_{S}$ respectively  account
 for the excitation by the Reynolds stress  and for the 
excitation resulting  from the advection of the  entropy fluctuations 
by the turbulent velocity field (the so-called entropy source term).
Here the entropy term ($F_S$) is an advective term
 which mixes turbulent pressure and entropy fluctuations.  
Expressions for $F_{R}$,  $F_{S}$ are :
\eqn{
\begin{array}{lll}
F_{R}  =   f_R(\xi) \,  S_R(\omega_0) &  &F_{S}  =   f_S(\xi)\,   S_S(\omega_0)  
\label{eqn:FR:FS}
\end{array}
\label{eqn:FS} \label{eqn:FR} 
}
with 
\eqna{
f_{R}(\xi) & = & {\cal G} \, \frac{\Phi}{3} \, 
\left (\deriv { \xi_r} {r} \right )^2 \, \label{eqn:fR} \\ 
f_{S}(\xi) & = &  {\cal H}  \,  
\left ( \frac{\alpha_s \, \tilde s}{\rho_0 \, w} \right )^2  \,   \frac{g_{\rm r}(\xi_{\rm r},m)  }{\omega_0^2}  \label{eqn:fS} 
}
where   $\displaystyle{\alpha_s =\left ( \partial p /\partial s  \right )_\rho}$,
$p$ denotes the  pressure and $s$ the entropy,
 $\tilde s$ is the rms value of the entropy fluctuations, and
 ${\cal G}$ and ${\cal H}$ are anisotropy  factors.
We assume that injection of acoustic energy 
into the modes is isotropic. This assumption 
implies ${\cal G}=16/15$ and ${\cal H}=4/3$ in 
Eq.(\ref{eqn:fR}) and Eq.(\ref{eqn:fS}) above.
Effects of the space averaged anisotropy 
in the driving process is investigated 
 in Sect.~\ref{sec:consequences:Velocity anisotropy at large scales}. 

The function $g_r(\xi_{\rm r},m)$  is defined as :
\eqn{
 g_{\rm r}(\xi_{\rm r},m) = 
\left( {1 \over\alpha_s } \deriv{\alpha_s }{r} 
  \,  \deriv { \xi_r} {r}  - \deriv{^2 \xi_r } {r^2} \right )^2
}

One can show that  the Reynolds contribution -~$P_R$~- scales 
as $\Phi^2 \, w^4$ while the source term involving 
the entropy fluctuations -~$P_S$~-  scales as $\Phi \,  w^2 \, \tilde s^2$.
$P(\omega)$ is thus very dependent of the estimated values 
of $w^2$, $\tilde s^2$ and $\Phi$.
The MLT provides  estimates for $w$ but $\Phi$ is a free parameter.
For isotropic turbulence $\Phi=3$, 
and in \citet[][BV-MLT hereafter]{Bohm58} 
formulation $\Phi=2$. In the present paper, unless otherwise stated,  
 $\Phi$ is   given by a simulation of the upper part of the solar
 convective zone  in Sect.~\ref{sec:Turbulent spectra}  below.

\vskip 0.3 truecm
For the driving sources in Eq.~(\ref{eqn:FR:FS})~:\\
\eqna{
S_{R} & = & \int_0^\infty { {\rm d} k \over  k^2} \,
  \frac{E(k,r)}{u_0^2} \, \frac{E(k,r)}{u_0^2} \, \chi_k ( \omega_0 ) \label{eqn:SR} \\
S_{S} & = &  \int_0^\infty { {\rm d} k \over  k^2} \,
  \frac{E(k,r)}{u_0^2} \,  \frac{E_s(k,r)}{\tilde s^2} 
\, \nonumber \\ & & \times \,\int_{-\infty}^{+\infty}{\rm d} 
\omega \, \chi_k ( \omega_0 + \omega)\chi_k (\omega) \; .\label{eqn:SS}
}

$E(k)$ represents the kinetic energy spectrum associated with the 
turbulent velocity field and $E_{\rm s}(k)$ models the spectrum of 
the turbulent entropy fluctuations, with $k$  the eddy wavenumber.
 The time-dependent part of the turbulent spectrum is described by the
function $\chi_k(\omega)$ which models the correlation time-scale of an 
eddy with wavenumber $k$. 
 The quantity $u_0 \equiv \sqrt{ \Phi/3}   \,  w $ 
is introduced for convenience (see Eq.~\ref{eqn:u0:w}),

The above expression for $P$ is mainly
 based on the assumption that the medium is 
incompressible. In other words,
 we adopt the Boussinesq approximation i.e. 
assume a homogeneous model for 
the turbulence and the excitation mechanism. 
We therefore neglect effects of the stratification in the excitation process.

\subsection{The turbulence model}
\label{The turbulence model}

 Let  $k_0(r)$ be the wavenumber at  which energy 
is injected into the turbulent cascade  and the energy $E(k)$  is maximum.
$k_0(r)$  is related  to the mixing-length 
$\Lambda \equiv \alpha \, H_{\rm p}$ by (Paper~I):
\eqn{
k_0^{\rm MLT}(r)\equiv \frac{2\pi}{\beta\,\Lambda(r)}=
\frac{2\pi}{\beta\,\alpha\,H_{\rm p}(r)}\,,
\label{eqn:k_0:MLT}
}
where $\beta$ is a parameter of order unity, $\alpha$ is the mixing-length
parameter and $H_{\rm p}$ is the pressure scale-height. 
This is a natural way to estimate $k_0$ as $\Lambda$ 
is the characteristic length of the largest convective elements.

The gaussian function is usually assumed for modeling
 $\chi_k(\omega)$ \citep[e.g.][]{Stein67,GK77} 
as a consequence of the turbulent nature of the
 medium where  the stochastic excitation occurs. 
The gaussian function  takes the form
\eqn{
\chi_k (\omega ) = \inv  { \omega_k \, \sqrt{\pi}}  e^{-(\omega / \omega_k)^2} \; .
\label{eqn:delta:omega}
}
where $\omega_k$ is its linewidth.

Let  $\tau_k$ be the characteristic time correlation length of an eddy  of  wavenumber $k$.
Eq.~(\ref{eqn:delta:omega}) corresponds  in  the time domain to  a gaussian function
with  linewidth equal to  $2 / \omega_k$.
Then  $\omega_k= {2  /  \tau_k}$ for a gaussian time spectrum.

The energy supply rate $P$ crucially depends on the correlation time-scale 
$\tau_k$ (see Paper~II). Following \citet{Balmforth92c} we define it as~:
\eqn{
\tau_k= \lambda \, (k \, u_k)^{-1}\,,
\label{eqn:tauk}
}
where $u_k$ is the velocity of an eddy with wavenumber $k$. 
The velocity $u_k$ is obtained from the kinetic energy spectrum $E(k)$ 
\citep{Stein67}
\eqn{
u_k^2 =  \int_k^{2 k}\,{\rm d}k\,E(k)\,.
\label{eqn:uk2}
}

$E(k)$ is normalised such that:
\eqn{
\int_0^{\infty}{\rm d}k\,E(k)  =  \ds  {1 \over 2} \,\overline{<\vec u^2> - <\vec u>^2 } \equiv   \ds {3 \over 2} \,  u_0^2(z) 
\label{eqn:E:normalisation}
}
where $u_0$ is introduced for convenience.
According to Eq.~(\ref{eqn:phiz}) and (\ref{eqn:E:normalisation}),  
$u_0$ and $w$ are then related to  each other by
\eqn{
{3 \over 2 } \, u_0^2 = {1 \over 2} \,   \Phi(z)   \, w^2(z)
\label{eqn:u0:w}
}

The parameter $\lambda$ in Eq.~(\ref{eqn:tauk}) 
accounts for our lack of precise  knowledge of the
 time correlation  $\tau_k$ in stellar conditions.
In the present paper, we assume $\lambda=1$ while $\beta$ (Eq.~\ref{eqn:k_0:MLT}) 
and $\Phi(z)$ (Eq.~\ref{eqn:phiz}) are   given by a simulation of the upper part of the solar convective zone
 in Sect.~\ref{sec:Turbulent spectra}  below.

\subsection{Computations of the excitation rate $P(\omega)$}

In practice, we compute the excitation rate $P(\omega)$ 
according to   Eq.~(\ref{eqn:P}). The calculation requires the knowledge of
several quantities which can be obtained either from a 1D model (Paper~II) 
or at least partly from a 3D simulation.
 Comparison of the results using both options yields insights in the excitation
 mechanism and its modelling. Hence in the following:

$\bullet$ The velocity, entropy fluctuations, anisotropy and turbulent spectra
$E's$   are obtained from a 3D
simulation as described in the next section. 

$\bullet$ The mean density,  $\rho_0$, the thermodynamic quantity 
$\alpha_s$, the oscillation properties -~eigenfrequencies 
and eigenfunctions~- are  calculated from  a
 solar envelope  equilibrium model and  \citet{Balmforth92a}'s pulsation code.  
The envelope  model is  built with a treatment of convection as prescribed 
by the  GMLT formulation and 
 is computed in the manner of
\citet{Balmforth92a} and \citet{Houdek99}. This solar envelope model 
 (hereafter GMLT solar model) is identical to the one 
considered in \citep[][hereafter]{Samadi00III}. In particular, it incorporates
turbulent pressure (momentum flux) in the equilibrium model envelope. 
The entire envelope is integrated using the equations appropriate to 
the nonlocal mixing-length formulation by Gough (1976) and to the Eddington 
approximation to radiative transfer \citep{Unno66}. 
The equation of state included a detailed treatment of the ionization of C, N, and O, and a treatment of the ionization of the next seven most abundant elements \citep{JCD82}, as well as 'pressure ionization' by the method of Eggleton, Faulkner \& Flannery \citep{Eggleton73}. 
In this generalization of the mixing-length approach, two  additional
 parameters, namely $a$ and $b$, are introduced which control 
the spatial coherence of the ensemble  of eddies contributing 
to the total heat and momentum fluxes ($a$), and the 
degree to which the turbulent fluxes are coupled to the local 
stratification ($b$). 
These convection parameters are calibrated to a solar model to obtain the 
helioseismically inferred depth of the solar convection zone of 0.287 of 
the solar radius (Christensen-Dalsgaard, Gough \& Thompson 1991).
The  adopted value for the shape factor  $\Phi=1.3745$,
a value which provides the best fit between computed solar damping rates 
and measurements by \citet{Chaplin98} \citep[see ][]{Houdek01}.
The detailed equations describing the equilibrium and pulsation models were 
discussed by \citet{Balmforth92a} and by \citet{Houdek96}.

\vskip 0.5 truecm

For implementation in Eqs.(\ref{eqn:P}-\ref{eqn:SS}), 
the quantities from the 3D simulation are interpolated at
 the GMLT model  mesh points. The grid of 
mesh points of the simulated domain is matched 
with the GMLT one such 
 that  $w$ in the 3D simulation 
has its maximum at the same layer as in the  GMLT model.
In the simulation, $w$ peaks $\sim 40$~km above 
the layer at  which the mean optical depth $<\tau>$  
is  unity while in the  GMLT model,
$w$ peaks $\sim 130$~km below the photosphere ($<\tau>= 2/3$).

\section{Constraints from the 3D simulation}
\label{sec:Constraints from the 3D simulation}


We consider a 3D simulation of the upper part of 
the solar convective zone obtained with the 3D 
numerical code developed at the Niels Bohr Institute for Astronomy, 
Physics and Geophysics (Copenhagen, Denmark).

The simulated domain is 3.2 Mm deep and its 
surface is 6 x 6 ${\rm Mm}^2$. The grid of 
mesh points is 256 x 256 x 163, the total duration 27 mn 
and the sampling time 30s. Physical assumptions are described in \citet{Stein98}.

Output of the simulation considered here are 
 the velocity field  $\vec u(x,y,z,t)$ and the entropy $s(x,y,z,t)$. 
They are used to determine the quantities 
  $\tilde s^2$, $\Phi(z)$, $w(z)$, $E_s(k,z)$, $E(k,z)$
which enter the excitation rate through  Eqs.~(\ref{eqn:P_RS},\ref{eqn:fR},\ref{eqn:fS},\ref{eqn:SR},\ref{eqn:SS}).

\subsection{Fourier transforms and averaging}
\label{sec:Data analysis}

We  compute the 2D Fourier transform, along horizontal planes,
of the velocity field $\vec u$ and the entropy $s$,
at each layer $z$. This provides 
$\vec {\hat u}(\vec k,z,t)$ and $\hat s(\vec k,z,t)$ where  $\vec k$ is the
wavenumber along the horizontal plane.
Next we integrate $\vec {\hat u}^2(\vec k,z,t)$ and  
 ${\hat s}^2(\vec k,z,t)$ over circles 
with radius $k$ at each given layer $z$.
Finally take a time average of the various
quantities over the time series. 
This yields $\vec {\hat u}(k,z)$ 
and ${\hat s}(k,z)$ where $k = \| \vec k \| $ is the wavenumber norm.

We define the time  averaged kinetic energy spectrum  $E(k,z)$ as:
\eqn{
E(k,z)  = \left \{ 
\begin{array}{lcl}
{ \ds 1 \over 2} \, \hat {\vec u}^2(k,z) & \textrm{for}  & k >0\\
\ds 0 &\textrm{for}  & k=0
\end{array}
\right . 
\label{eqn:def:E}
}
and the time averaged  spectrum of the entropy  $E_s(k,z)$ as:
\eqn{
E_s(k,z)  = \left \{ 
\begin{array}{lcl}
{ \ds 1 \over 2} \, \hat {s}^2(k,z) & \textrm{for}  & k >0\\
\ds 0 &\textrm{for}  & k=0
\end{array}
\right . 
\label{eqn:def:Es}
}

From Parseval-Plancherel's relation,  $E(k,z)$  and $E_s(k,z)$  satisfy:
\eqn{
\begin{array}{lllll}
\vspace{0.2cm}
\ds \int_0^{+\infty}  {\rm d} k  \, E(k,z) & =  & \ds  {1 \over 2} \, \overline{<\vec u^2> - <\vec u>^2 }& \equiv  & \ds {3 \over 2} \,  u_0^2(z) 
\label{eqn:norm:E}
 \\
\vspace{0.2cm}
\ds \int_0^{+\infty}  {\rm d} k  \, E_s(k,z)   & =  & \ds {1 \over 2} \, \overline{ <s^2> -  <s>^2 } & \equiv  &\ds {1 \over 2} \, \tilde  s^2(z)
\end{array}
\label{eqn:norm:Es}
}
Hence the definitions of the energy spectra here  involve zero mean velocity
and entropy fluctuations.

\subsection{Convective velocities and entropy fluctuations}
\label{sec:Convective velocities and entropy fluctuations}

\fig{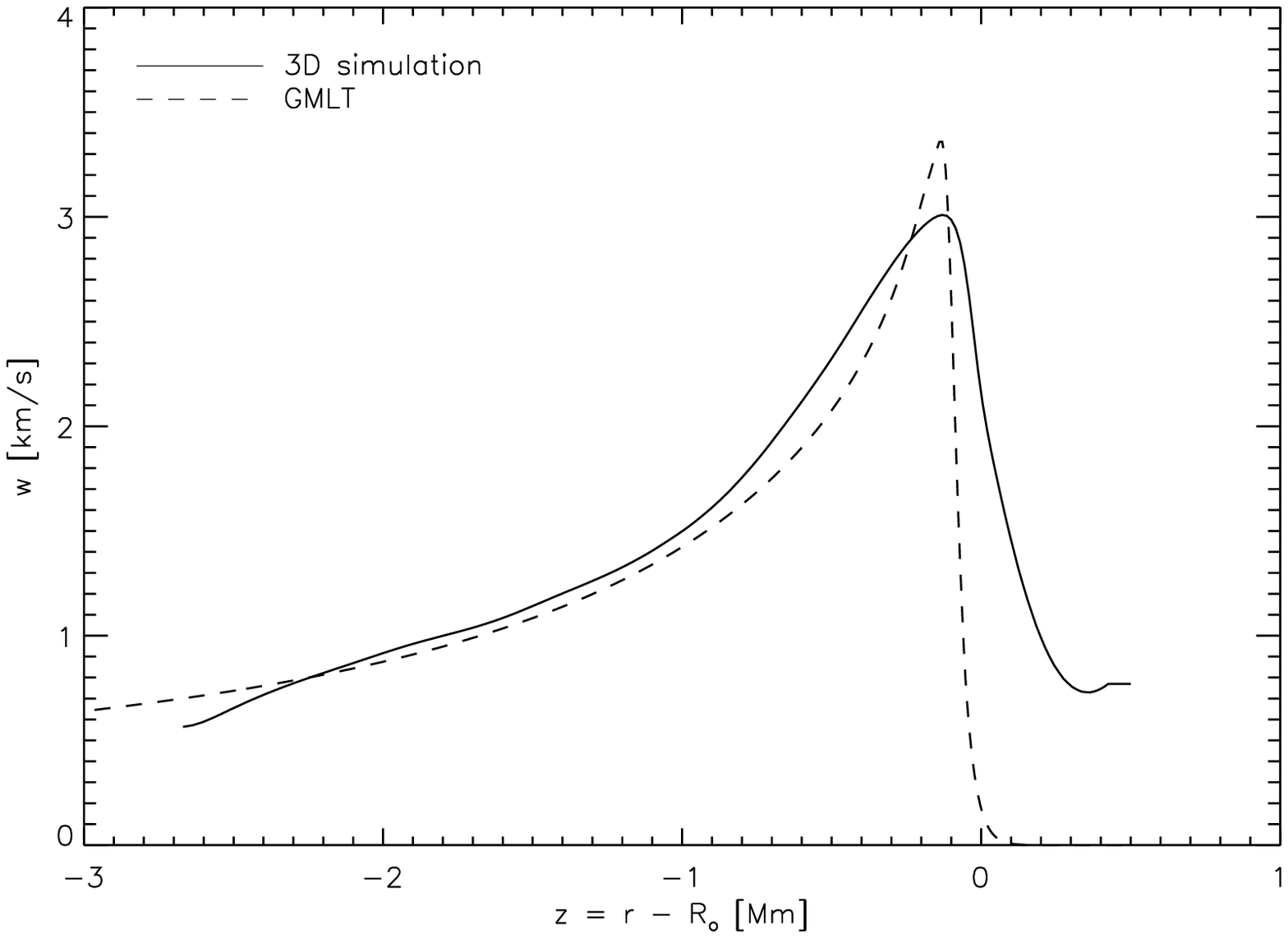}{The root mean square
 of the vertical component of the velocity 
($w=\sqrt{<u_z^2>-<u_z>^2}$) in the upper layers of a solar model  
is plotted versus  depth   for the 3D simulation 
(solid line)  and for  the 1D GMLT model 
 (dashed line). The abscissa is the  depth $z = r -R_\odot$ 
where $R_\odot$ is the radius at the photosphere.
The $w$ maximum  corresponds to the top of 
the superadiabatic region and  is reached at the 
depth $z \simeq - 130~{\rm km}$ in the GMLT model. 
The grid of mesh points of the simulated domain is 
adjusted  so that the $w$ maxima of the 3D simulation and the GMLT coincide at 
the same layer. 
}{fig:w_z}

\fig{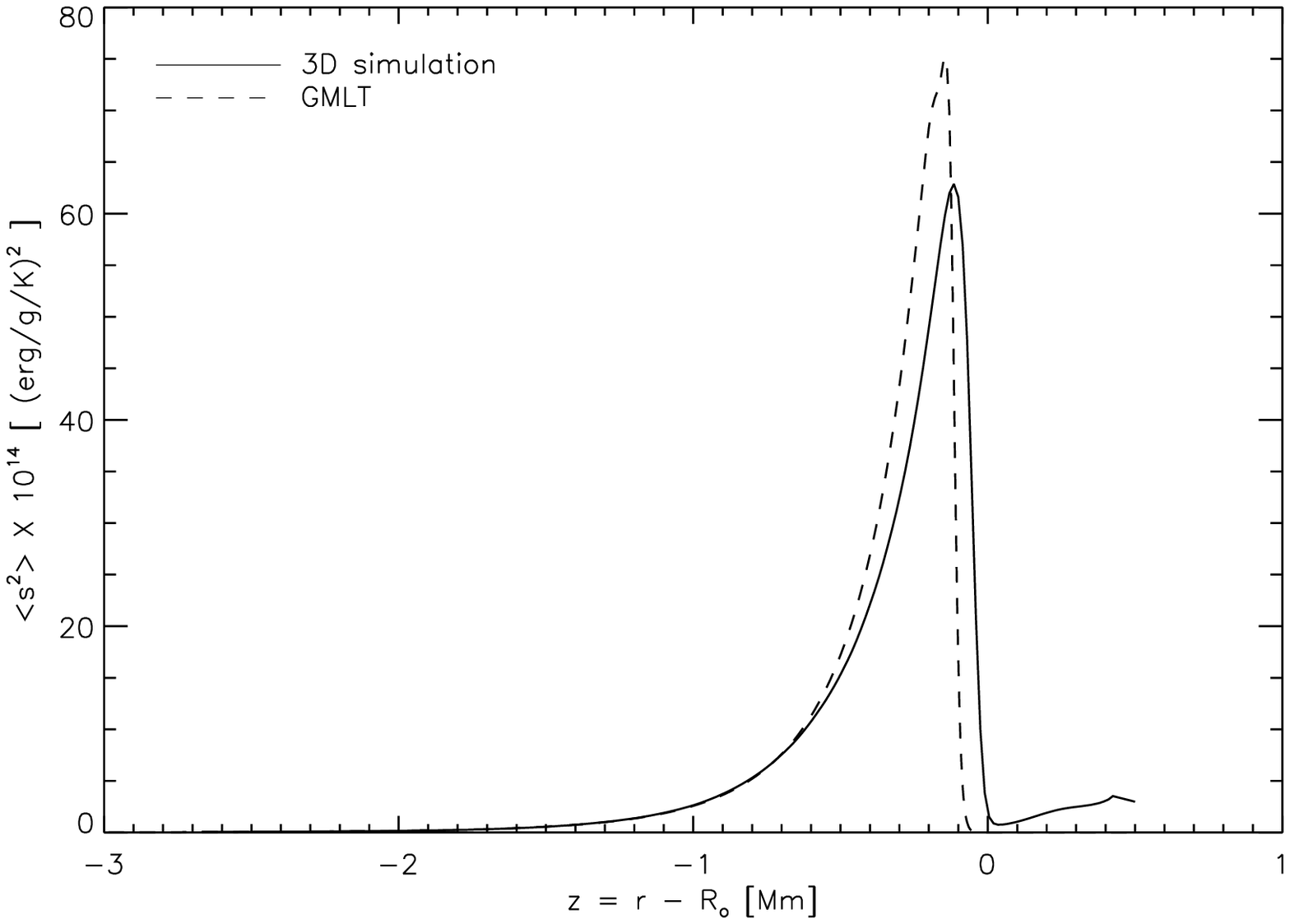}{Same as Fig.~\ref{fig:w_z} for the 
mean square of the entropy fluctuations ($\tilde s^2$).
 The peak is narrower than for the velocity $w$   because $\tilde s^2$ 
scales approximatively as $w^4$.}{fig:ss_z}

\fig{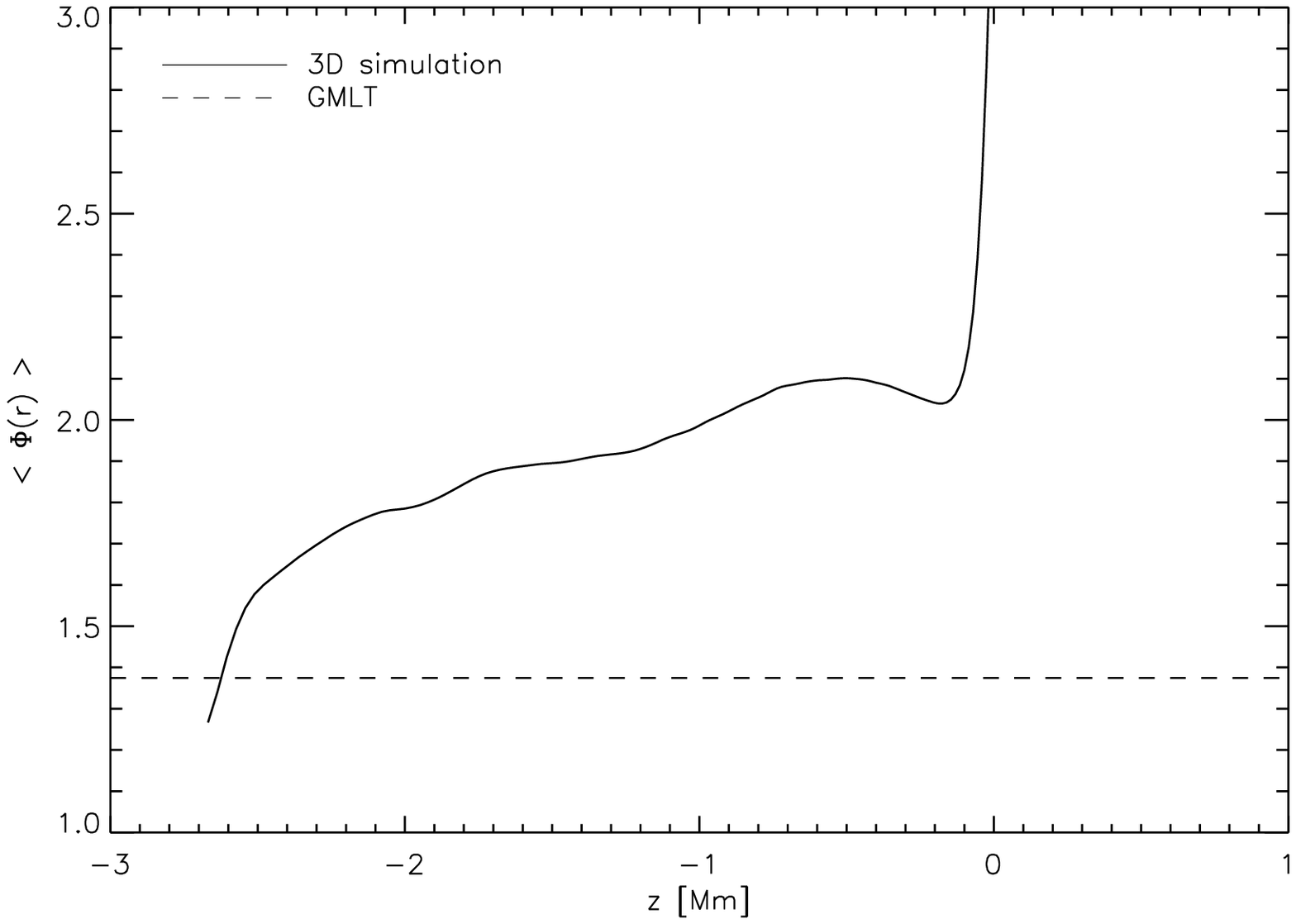}{Same as Fig.~\ref{fig:w_z} 
for  the anisotropy factor $\Phi$ versus $z$.}{fig:phi_z}

Figs.~\ref{fig:w_z},~\ref{fig:ss_z}  present $w(z)$,
$\tilde s^2 (z)$   versus depth for the 3D simulation.  
For comparison purpose, the plots also 
show $w$ and $\tilde s^2$ obtained 
with  the GMLT solar model.

The vertical velocity GMLT  $w$ is larger at the top of 
superadiabatic region but smaller just  
beneath compared to values from the simulation.  
The GMLT  $\tilde s^2$  is  larger than in
the simulation ($\sim 20$~\%). This explains that 
the  relative contribution  of the  entropy source term 
to the excitation  is overestimated  
with the GMLT model 
(see Sect.~\ref{sec:consequences:Convective velocities and entropy fluctuation}).
 Differences between
the GMLT and the simulation are likely to be related to differences in
the convective efficiency: GMLT is less efficient than the 3D.
Indeed as pointed out by \citet{Houdek02}, 
a single eddy approach such as the GMLT results in a larger 
peak for the superadiabatic gradient.

\subsection{Velocity  anisotropy at large scale}
\label{sec:Large scale anisotropy}

As it will be shown in 
Sect.~\ref{sec:consequences:Velocity anisotropy at large scales}, 
the value of $\Phi$ plays a crucial role in controlling 
the depth of the excitation region and therefore the total
 amount of acoustic energy injected into the  oscillation modes.

Fig.~\ref{fig:phi_z} displays  the anisotropy factor
 $\Phi$  versus depth $z$ for the 3D simulation.  
$\Phi(z) $ sharply decreases from the value $\Phi=3$ at the top 
of the CZ down to $\Phi=2$ and then 
 slowly decreases to reach  the value $\Phi\simeq  1.3$ 
at the bottom of the simulation. 
The decrease of $\Phi(z) $ with depth 
is explained first by the onset of the convection 
and the formation of convective plumes 
at $z \sim 0$ and then by the \emph{relative} increase
in number  of the plumes  inward in the simulation.
Indeed, plumes are highly anisotropic structures  
whereas turbulent cells  are quite isotropic.
The turbulent Mach number increases with $z$ and 
reaches its maximum value at the top of the CZ.
Therefore the fluid is more turbulent  outward in the atmosphere.
Consequently the number of turbulent  isotropic cells increases 
with $z$ up to the top of the CZ whereas the number of plumes remains roughly constant. The medium is thus 
more isotropic outward than inward.

In most of the excitation region,
the value of $\Phi=2$ consistent with the BV-MLT
 is in better agreement with the values
 of $\Phi(z) $  inferred from the simulation
 compared to the value  $\Phi=1.3745$
which must be  imposed  for the  GMLT solar model
in order to match the observed solar damping rates.

\subsection{Turbulent kinetic energy spectrum $E(k)$}
\label{sec:Turbulent spectra}

\begin{figure}
        \resizebox{\hsize}{!}{\includegraphics  {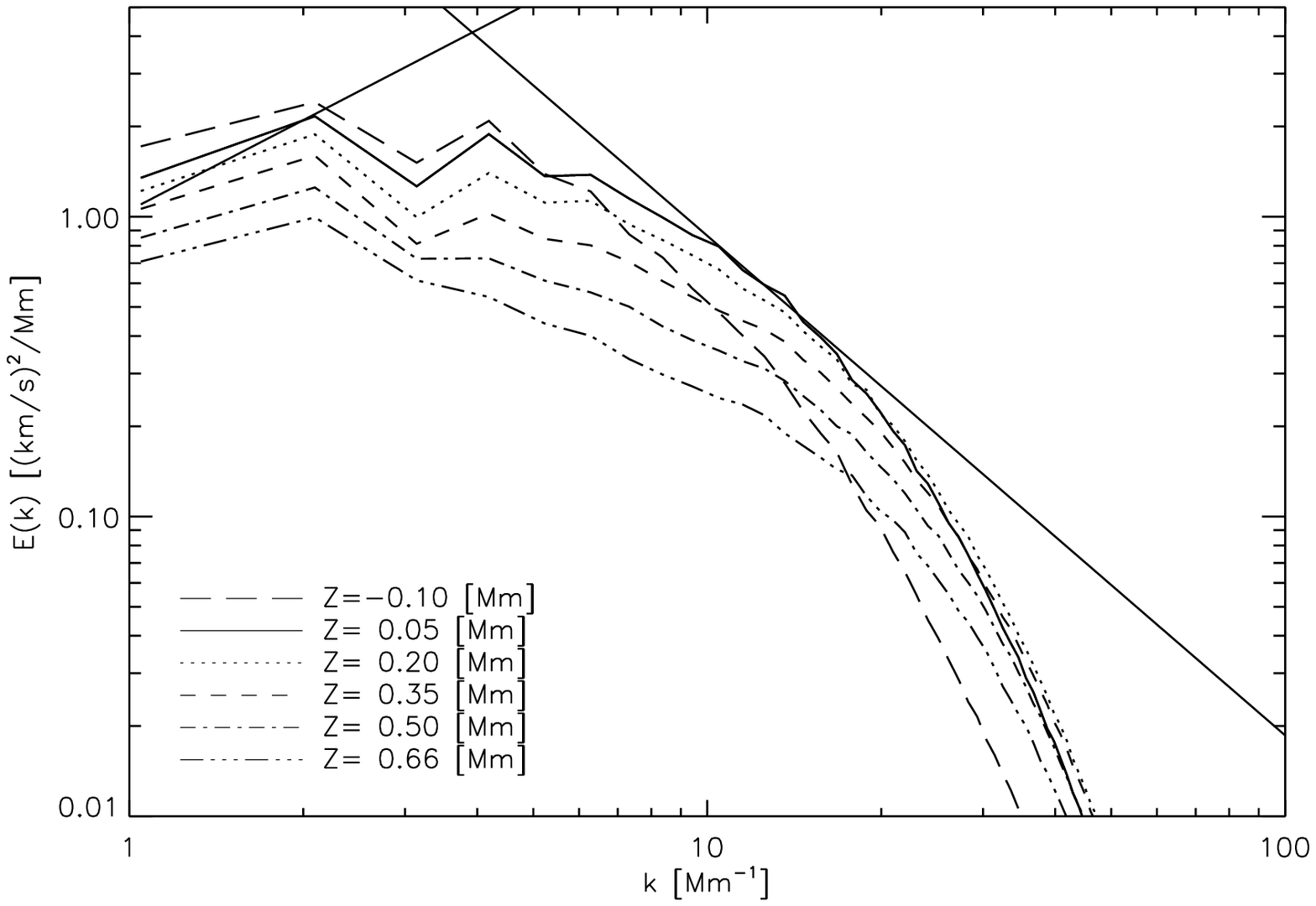}}
        \resizebox{\hsize}{!}{\includegraphics  {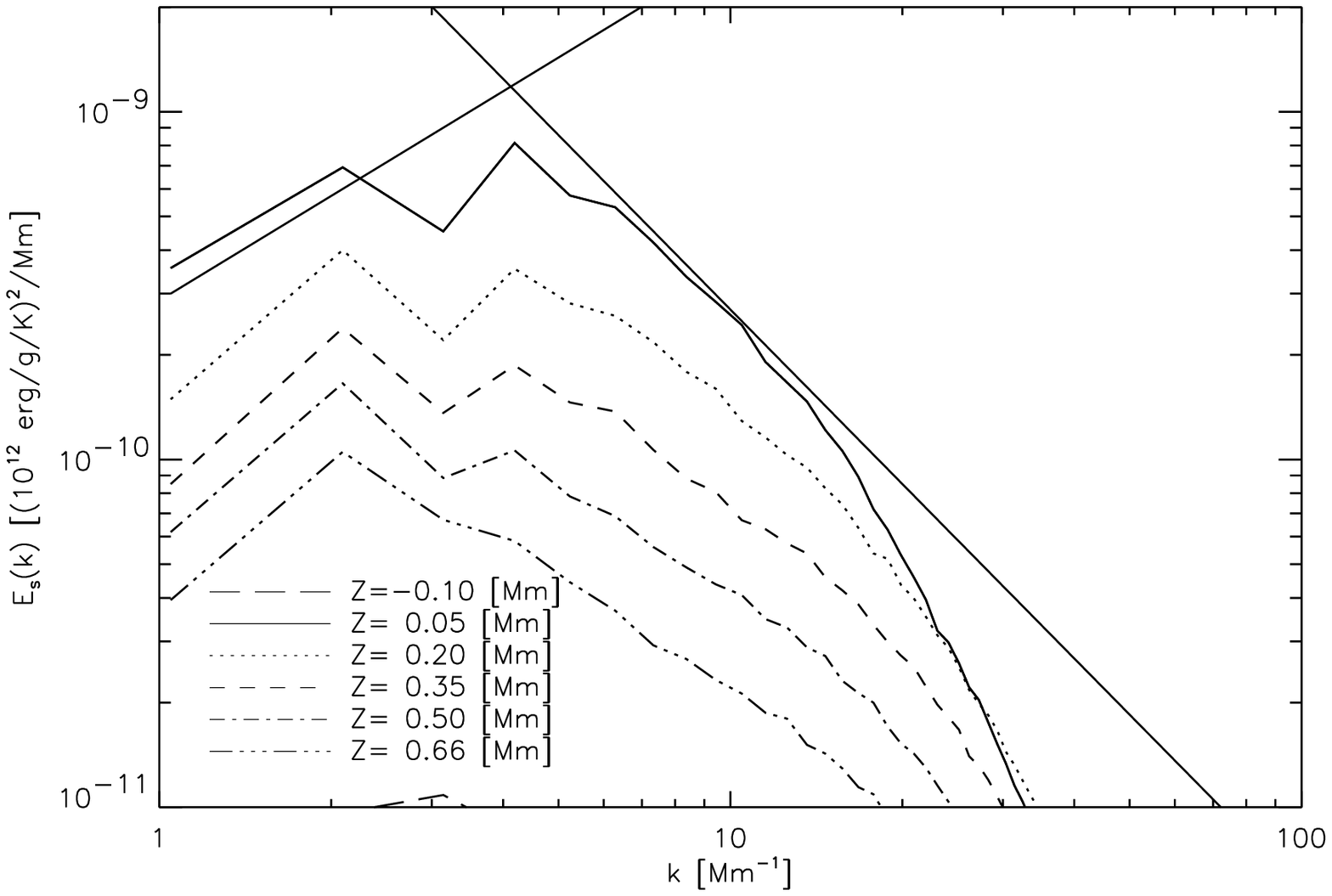}}

\caption{Turbulent kinetic energy spectra 
$E$ (top)  and  $E_s$ (bottom)  
from the simulation are plotted versus $k$ and 
for different depths  $z$ in the simulation. The straight solid lines
delimitate the slopes $k^{1}$ and $k^{-5/3}$ of the NKS spectrum. Intersection
of the slopes determines $k_0^E$, the scale of maximum energy  at each depth $z$ }
        \label{fig:EEzEs_z}
        \end{figure}

Variations of   $E$ and $E_s$ with $k$ at different depths 
$z$  are depicted in  Fig.~\ref{fig:EEzEs_z}.
The  spectra clearly show two regimes:
 at large scale (small values of $k$), the spectra 
increase approximately as $k^{+1}$ which can
 roughly be explained using dimensional analysis. 
At small scale (large values of $k$), the spectra 
decrease very rapidly with $k$.
The Kolmogorov law ($k ^{-5/3}$) is observed only
over a small $k$-range.  
Departures of the computed spectra  from a  Kolmogorov law  
 at high values of $k$ can be explained by  the   finite resolution 
of the simulation spatial grid.

\vskip 0.3 truecm
The main characteristics of the 
  kinetic spectrum $E(k,z)$ -~$k$ dependency~-
 derived from the 3D simulation are approximatively  reproduced
 by an analytical expression which was considered by \citet{Musielak94}, namely 
the 'Extended Kolmogorov Spectrum' (EKS hereafter) 
defined in \citet{Musielak94} as:
\eqn{
 E(k,z) = a \, \frac{ u_o^2}{ k_0^E } \, \left \{
\begin{array}{lcl}
\vspace{0.2cm}
\ds \left ( \frac{k}{k_0^E} \right )^{+1} & \textrm{for} & k < k_0^E (z) \\
\ds \left ( \frac{k}{k_0^E} \right )^{-5/3} & \textrm{for} & k > k_0^E (z) 
\end{array}
\right .
\label{eqn:EKS}
}
where $a$ is a normalisation factor which satisfies Eq.~(\ref{eqn:norm:E}) 
and $u_0$ is  defined according to Eq.~(\ref{eqn:norm:E}). $k_0^E$ is the scale
of maximum energy in the energy spectrum.

At each layer, $k_0^E$ is determined 
by imposing that the EKS, as defined above, 
 matches the turbulent spectrum  $E(k,z)$ calculated 
from the simulation as well as possible.
This then fixes  the $z$ dependency 
of $k_0^E$. A similar procedure is applied 
for $E_s(k,z)$ for which we introduce $ k_0^{E_s}$.
All  spectra satisfy their respective normalisation
condition as given in Eq.~(\ref{eqn:norm:E}).

For comparison, in Fig.~\ref{spc_cinetique},   the
`Nesis Kolmogorov Spectrum'' (NKS hereafter) determined
 from  solar observations of  \citet{Nesis93} is also shown.
The NKS scales as  $k^{-5}$  {\it in the energy injection region}  
for  $k < k_0^E$ and down to $k_{min}=0.7\,k_0^E$. 
This spectrum does not agree with turbulent spectrum  $E(k,z)$
 calculated from the simulation.
In particular, the NKS underestimates the 
velocity of the small size turbulent elements 
in the cascade  ($k>k_0$) and 
overestimates the velocity of the turbulent 
with wavenumber $k \sim k_0^E$.
As we will show in Sect.~\ref{sec:consequences:Turbulent spectra}, differences
between the EKS and the NKS have an important impact on $P(\omega)$.

If we assume that $\ds k_0^{E} =  k_0^{E_s}$, 
one can show that $P(\omega)$  scales as $k_0^{-4}$. 
$P(\omega)$ is therefore very dependent on the values 
reached by $k_0(z)$ in the excitation region.
Variation of $k_0(z)$ with depth is thus shown in Fig.~\ref{fig:k0_z} 
for  $E$ and $E_s$: 
$\ds k_0^{E_s}$ and $\ds k_0^{E}$ vary slowly within the excitation region.

For comparison, in Fig.~\ref{fig:k0_z} we have also
plotted $\ds k_0^{\rm MLT}(z)$, the
MLT value for  $k_0(z)$ according to
Eq.~(\ref{eqn:k_0:MLT}). 
The scaling parameter $\beta$ in the definition of 
$k_0^{\rm MLT}$ is determined such that 
 $k_0^{\rm MLT}$ and $k_0^{\rm E}$ take the same value 
  at  the layer $z \simeq -130~{\rm km}$ where $w$ reaches 
its maximum (and consequently the layer where the excitation is maximum).
The derived value  is $\beta=3.48$.

\fig{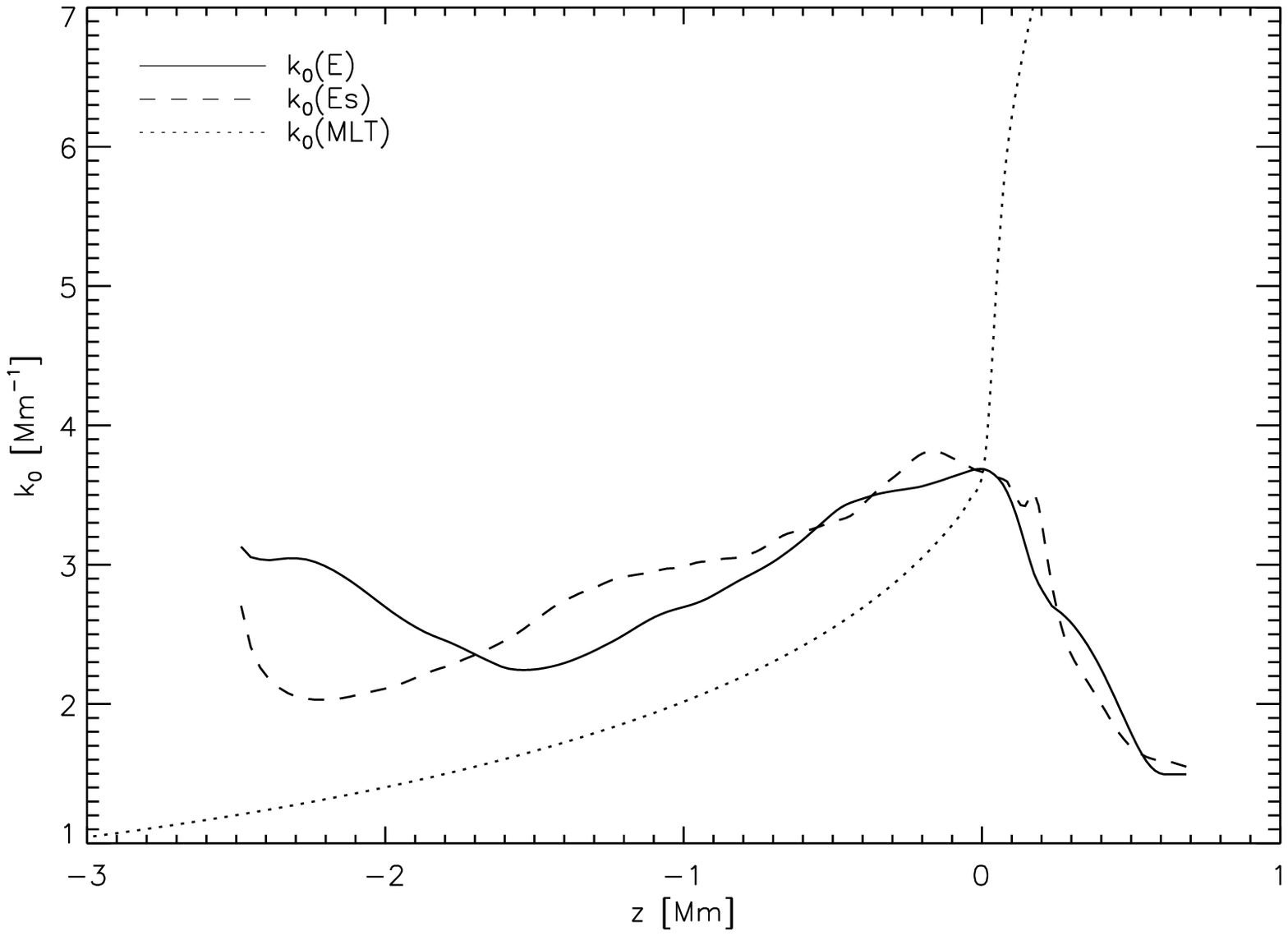}{The wavenumbers $k_0^E$ (solid line)  and $ k_0^{E_s}$ (dashed line) 
are plotted versus $z$ ($k_0^E$ and $ k_0^{E_s}$  are obtained   by 
fitting, at each layer,  the   EKS (Eq.~\ref{eqn:EKS}) 
 to   the computed spectrum  $E$ and $E_s$ of  Fig.~\ref{fig:EEzEs_z} resp., see text for details).
The dotted line corresponds to $k_0^{\rm MLT}(z)$ obtained
 according to Eq.~(\ref{eqn:k_0:MLT}). In computing $k_0^{\rm MLT}(z)$,
we assume $\beta=3.48$ in order for $k_0^{\rm MLT}$ to match 
the value reached by $\ds k_0^{E}$ (solid line) 
at the layer $z \simeq -130~{\rm km}$ where $w$ reaches
 its maximum ($k_0^{E}=3.62~{\rm Mm}^{-1}$ at that layer).} {fig:k0_z}

 $k_0^{\rm MLT}$ varies slowly with depth below the top of 
the superadiabatic region ($z \simeq -130~{\rm km}$) but 
increases very rapidly above. 
Such a  behavior is explained by the rapid decrease 
with $z$ of the pressure scale height $H_{p}$ 
(which enters in the definition of $k_0^{\rm MLT}$, Eq.~(\ref{eqn:k_0:MLT}))
 in the atmosphere. 

 
Comparison between  $k_0^{\rm MLT}(z)$ and  
$k_0^{E}(z)$ shows that the mixing-length 
approach does not model satisfactorily the 
behavior of  $k_0^{E}(z)$ in particular 
just above the layer at which $w$ reaches its maximum value. 
Consequences  in terms of mode excitation 
are investigated in Sect.~\ref{sec:consequences:stratification of the turbulent spectrum}.

\section{Consequences in term of p~modes excitation}
\label{sec:Consequences in term of p modes excitation}


The  acoustic energy supply rate $P$ 
injected into the solar oscillations is related to the rms value $v_s$ of  surface velocity as :
\eqn{
P (\omega_0) = 2 \eta \, \frac{I  }{ \xi_r^2(r_s) } \, v_s^2 (\omega_0) 
\label{eqn:P_vs2}
}
where  $\eta$ is the mode damping rate and
$r_s$ is the radius at which oscillations are measured.

We derive the  `observed' $P$ from \citet{Chaplin98}'s seismic 
data according to Eq.~(\ref{eqn:P_vs2}) where the mode damping rate ,$\eta$, and the mode surface velocity , $v_s$, are obtained from \citet{Chaplin98}'s data.  The mode mass ${I /  \xi_r^2(r_s)}$ is given by  
the GMLT model and we adopt  $r_s = R_\odot + 200$~km  
consistent with the   observations.

Theoretical values of   $P$ are computed according to  Eq.~(\ref{eqn:P}).
In Eqs.(\ref{eqn:SR}) and (\ref{eqn:SS}) 
the integrations over $k$ are performed
 from $k=k_{min}$ (where $k_{min}$ depends 
on the adopted turbulent spectra $E$ and $E_s$) 
to $k=20 \, k_0$. 
We checked   numerically that 
contributions  to the excitation rate from  turbulent elements
 with $k \gtrsim 20\,k_0$   
are negligible. 

A  gaussian function is assumed for $\chi_k(\omega)$ in Eq.~(\ref{eqn:SR}) and Eq.~(\ref{eqn:SS}). 

For the other quantities 
($w$, $\tilde s^2$, $\phi$, $k_0$, $E(k/k_0)$ and $E_s(k/k_0)$) involved in the
 expression for $P$ we investigate several possible assumptions.


\subsection{Convective velocities and large scale anisotropy}
\label{sec:consequences:Convection and large scales anisotropy}

In this section,  the excitation rate $P$ (Eq.~(\ref{eqn:P}))
 is computed with the following  assumptions: \\
-  the $k$-dependency $E$ and $E_s$ is given  by the analytical form
of Eq.~(\ref{eqn:EKS}), also called the EKS.\\
- $ k_0^{\ds E}= k_0^{\ds E_s} = k_0^{\rm MLT}$  
where  $ k_0^{\rm MLT}(z)$ is given in Eq.~(\ref{eqn:k_0:MLT})
with $\beta=3.48$  so that  $k_0^{\rm MLT}$ takes the 
value reached by $k_0^{E}(z)$ (solid line, Fig.~\ref{fig:k0_z}) 
at the layer $z \simeq -130~{\rm km}$  where $w$ reaches its maximum.

For the quantities $\Phi$, $w$ and $\tilde s^2$ 
we investigate the effects of using either 
the values derived from the 3D simulation (see Sect.~\ref{sec:Convective velocities and entropy fluctuations} and Sect.~\ref{sec:Large scale anisotropy})  or calculated 
with the GMLT solar model.

\subsubsection{Convective velocities and entropy fluctuations}
\label{sec:consequences:Convective velocities and entropy fluctuation}

The values of $w$, $\tilde s^2$ and $\Phi(z)$ 
are fixed by the 3D simulation inside the simulation
domain and by the  1D equilibrium model outside this domain.
 Either, if we impose zero values 
or if we assume quantities from the 1D MLT model,
no sensitivity on the calculation of $P$ is found.

Results are shown  in Fig.~\ref{Poscgh_MLT} for $P$ and for 
the relative contribution of the Reynolds stress 
 to the total energy supply rate $P$.  When the excitation rate $P(\nu)$ is
 computed with quantities derived from the 3D simulation as described in
 Sect.2.3, the resulting 
excitation rate at maximum is  found too small  by a factor $\sim 4$ 
compared with the observations.

Provided the appropriate value for $\Phi$ is given 
in the GMLT estimations (see Sect.4.1.2 below), 
no significant difference is found 
in the excitation rate when computed with  the values of $w$ and
 $\tilde s^2$  from the simulation 
or their respective  GMLT estimations.

The main effect is illustrated in the bottom panel of Fig.~\ref{Poscgh_MLT}:
the 3D simulation generates a
 larger  relative  contribution of the Reynolds stress 
to $P$  than the GMLT model. This is explained as follows:
within most part of the excitation region - except at the top of
superadiabatic region - the  values reached by $w$ are larger 
 whereas
values reached by $\tilde s^2$ 
 are smaller
 than their corresponding  GMLT estimations.

\subsubsection{Velocity anisotropy at large scales} 
\label{sec:consequences:Velocity anisotropy at large scales} 
\begin{figure}
\resizebox{\hsize}{!}{\includegraphics  {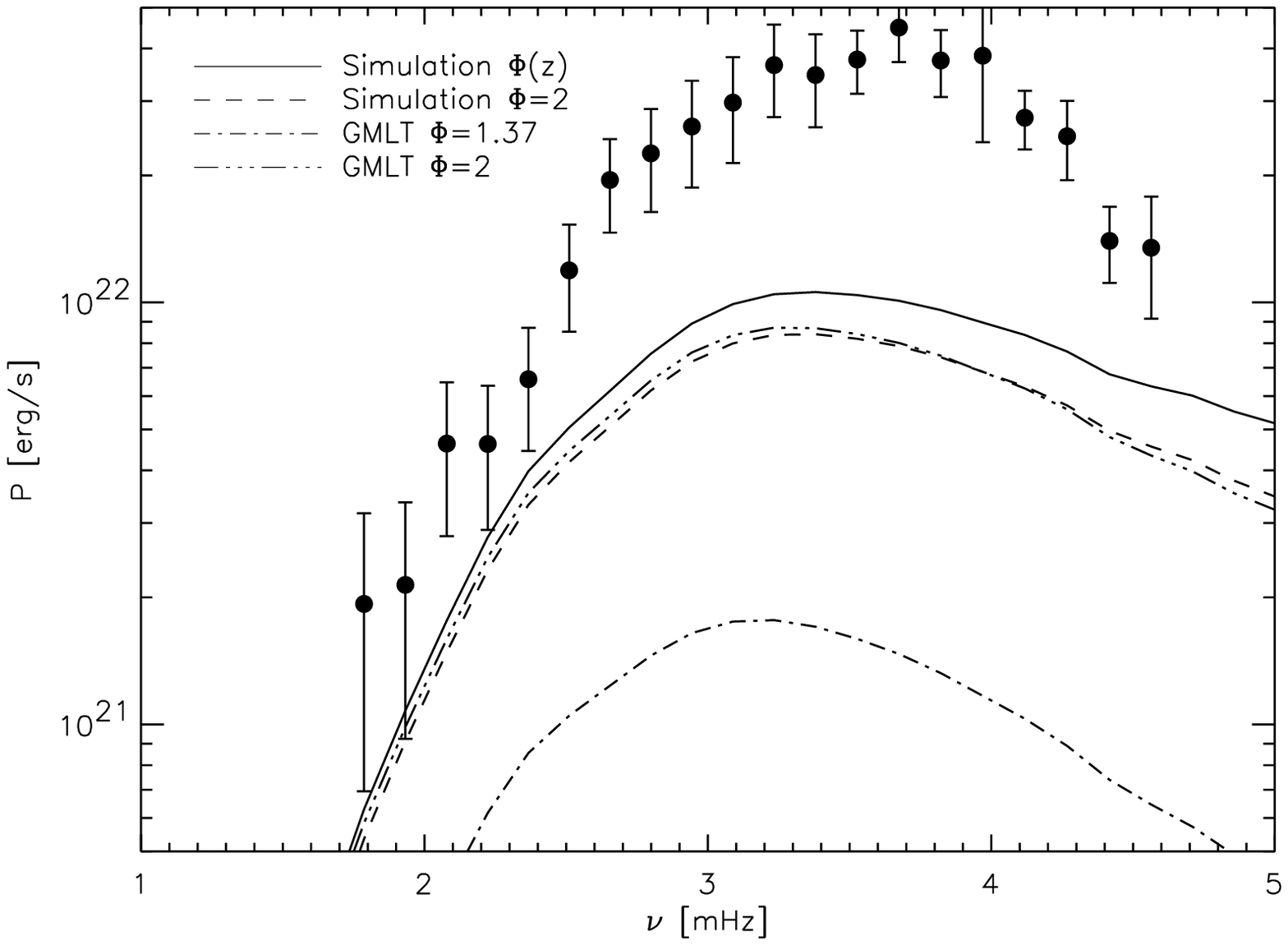}}
\resizebox{\hsize}{!}{\includegraphics  {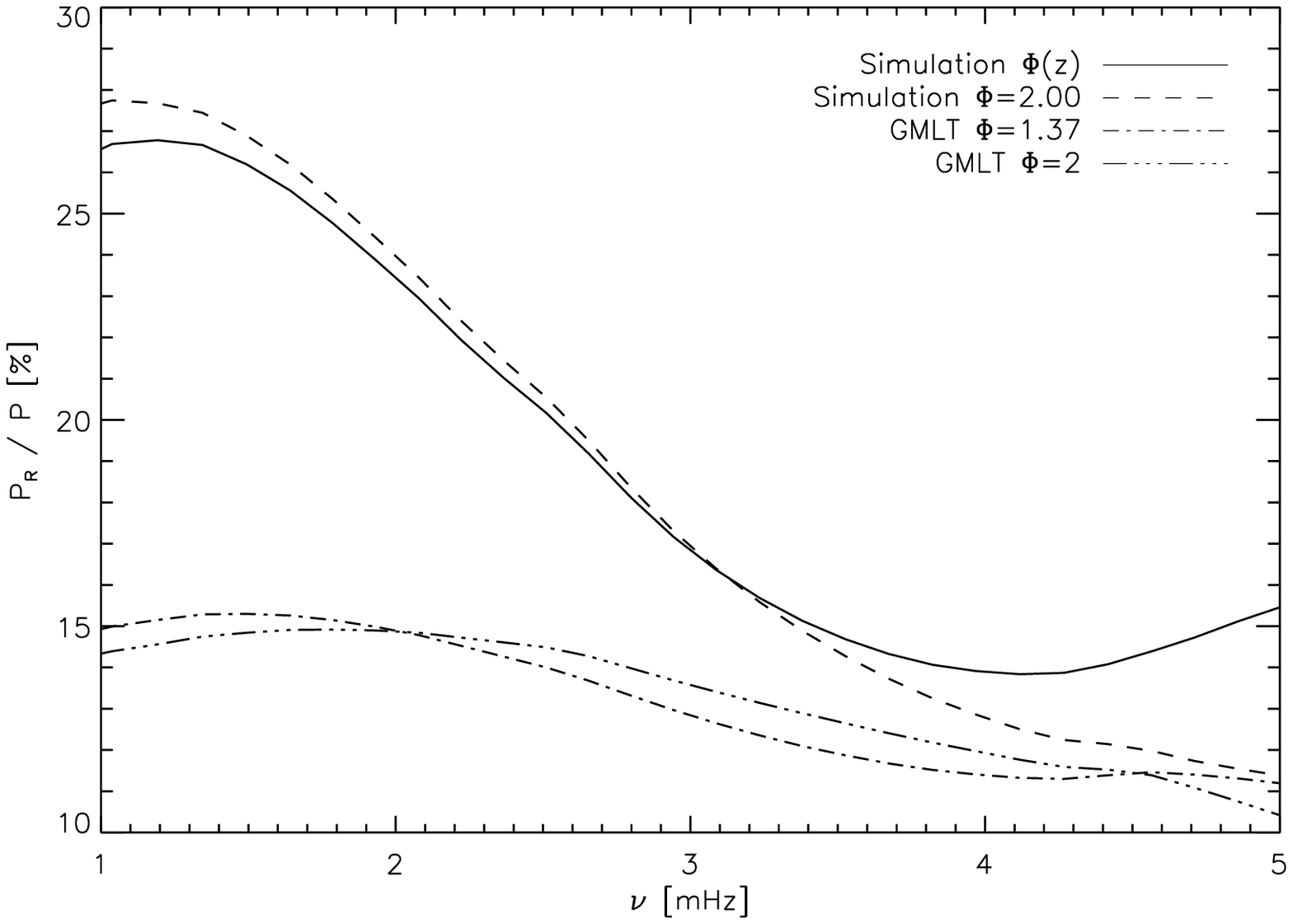}}
\caption{{\bf Top:} Rate $P$ at which acoustic energy 
is  injected into the solar radial modes. 
The filled dots represent $P$ computed from \citet{Chaplin98}'s 
solar seismic data according to Eq.~(\ref{eqn:P_vs2}).
The curves represent theoretical values of 
$P$ computed  according Eq.~(\ref{eqn:P}) 
and for different computations of $w$, $\tilde s^2$ and $\Phi(z)$:
Solid line: values of $w$, $\tilde s^2$ and $\Phi(z)$ 
are fixed by the 3D simulation inside the simulated 
domain and by the  1D equilibrium model outside this domain.
Dashed line ($-~-~-$): same as solid line but
 fixing $\Phi$ to BV's value $\Phi=2$.
Dot dashed line ($-~.~-~.-$) : values of 
$w$, $\tilde s^2$ and $\Phi~(=1.37)$ 
are fixed by the 1D equilibrium model (GMLT). 
Three dots dashed line  ($-~...~-~...~-$): same as 
the {\bf dot dashed line  but fixing $\Phi$ to BV's value $\Phi=2$.
 Bottom:} Same as top panel for the relative 
contribution of the Reynolds stress, $P_{R}$, 
to the total acoustic energy $P$. 
}
\label{Poscgh_MLT}
\end{figure}

The main consequence (in term of p~modes excitation)  of the
differences between the  time averaged properties of the convective region
inferred from the 3D simulation and from the GMLT solar model (Fig.~\ref{Poscgh_MLT})
is due to  differences in their respective anisotropy factor
 $\Phi$ values (Fig.~\ref{fig:phi_z}).

Within most of the excitation region, $\Phi(z)$ is found 
close to $\sim 2$ and thus larger than the value $\Phi=1.37$ 
assumed for the 1D equilibrium model 
(see Sect.~\ref{sec:Large scale anisotropy}
 and Fig.~\ref{fig:phi_z}).
Smaller values of $\Phi$ decrease 
the rms total convective velocity  which  results in larger values of 
$\tau_k$ (see Eq.~\ref{eqn:tauk}- Eq.~\ref{eqn:uk2}) 
and therefore in a smaller depth of excitation 
for a given mode frequency (see Paper~III for more details).
Smaller values of the rms total convective velocity 
also  induce  smaller values of 
$E$ in the integrand of Eq.~(\ref{eqn:P}). 
Consequently, as it is
 illustrated in Fig.~\ref{Poscgh_MLT}, 
the total amount of acoustic energy injected into the modes 
is $\sim 5$ times smaller for the constant 
value $\Phi=1.37$ compared to the constant 
value  $\Phi=2$ 
(the relative contribution of the Reynolds stress to $P$ is 
found $\sim 2$ times larger in the simulation).

The effect of the depth dependency of $\Phi$ on
 the mode excitation  is small except at high frequency. 
This is illustrated in the bottom panel 
of Fig.~\ref{Poscgh_MLT} (compare the solid line with the dashed line). 
Just above the top of the superadiabatic region
 ($z \gtrsim -130~$km), 
$\Phi(z)$ increases rapidly with $z$ 
until the value $\simeq 3$ (Fig.~\ref{fig:phi_z}).   Most of the 
injection of acoustic energy  into the high 
frequency modes occurs at the top of superadiabatic region. 
The high frequency modes are therefore 
 more sensitive to this rapid increase of $\Phi(z)$. As a 
consequence, the relative contribution 
of the Reynolds stress is larger for the 
high frequency modes than it is 
when assuming the constant value  $\Phi=2$.

\subsection{Turbulent spectra}
\label{sec:consequences:Turbulent spectra}

In this section we compare
the excitation rate obtained assuming, for the turbulent spectra 
($E$ and $E_s$) either -~the EKS spectrum (Eq.~\ref{eqn:EKS}) with slopes given by the 3D simulation 
as in the previous section-~ or  
assuming  the NKS spectrum  from  solar observations of  \citet{Nesis93}
(see also  Paper~I).

As in Sect.~\ref{sec:consequences:Convection and large scales anisotropy}
 we compute $P$ 
using  $w$, $\tilde s^2$ and $\Phi$  derived 
from the 3D simulation and assuming that $k_0^E= k_0^{E_s} = k_0^{\rm MLT}$.
The results are plotted in Fig.~\ref{Poscgh_synspc}.

The NKS overestimates the maximum in $P$ by a
 factor $\sim 1.5$ while the EKS underestimates it by a factor $\sim 4$. 
This is because most of  the kinetic energy in the 
NKS  is concentrated at $k \sim k_0^E$ whereas
 in the EKS a large part of the energy is concentrated  both at large scales ($k < k_0^E$) 
and at small scales ($k > k_0^E$).



\vspace{0.5cm}

\fig{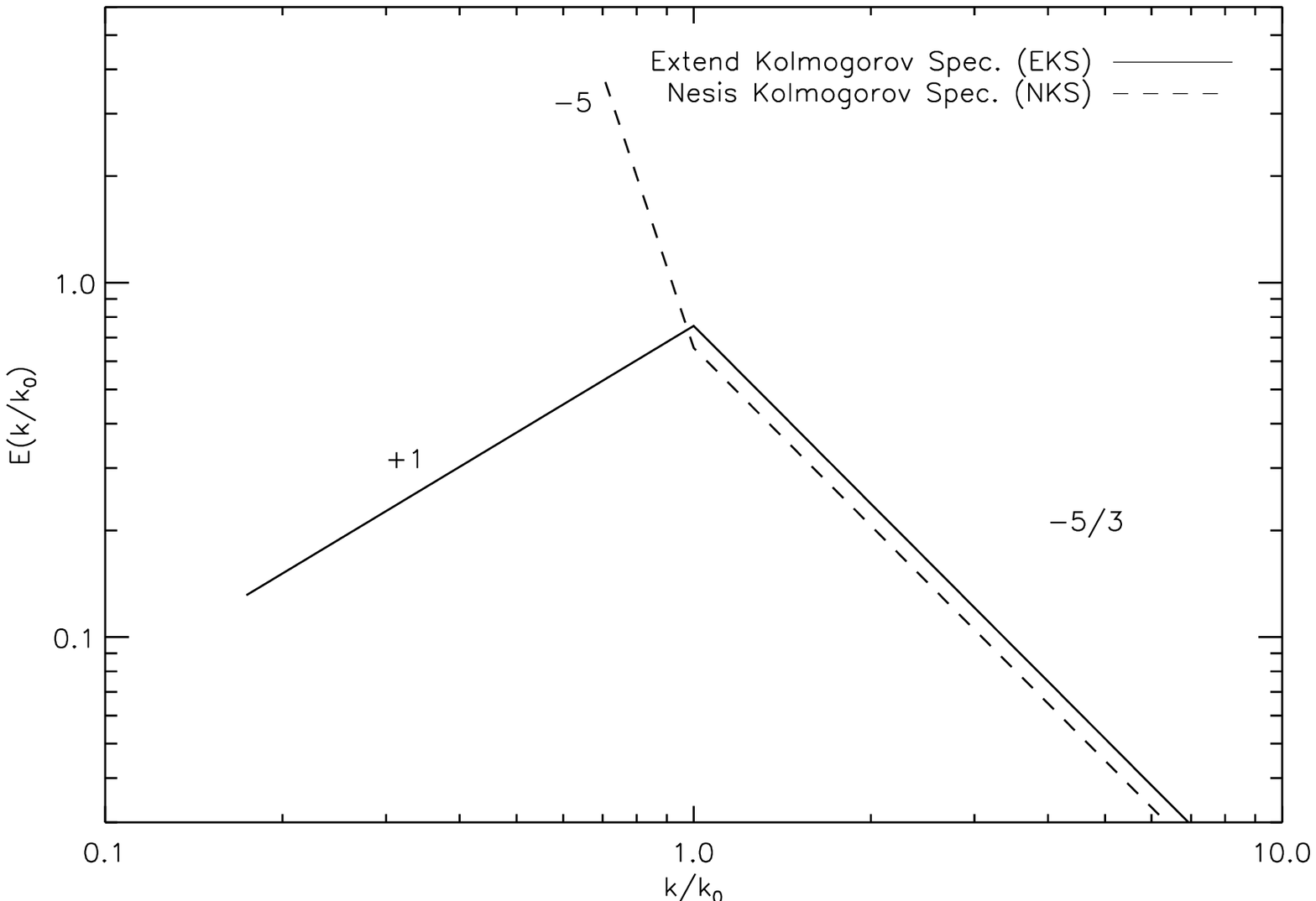}{The NKS and the EKS 
turbulent kinetic energy 
spectra are plotted versus the normalized 
wavenumber $k/k_0$.
}{spc_cinetique}
\fig{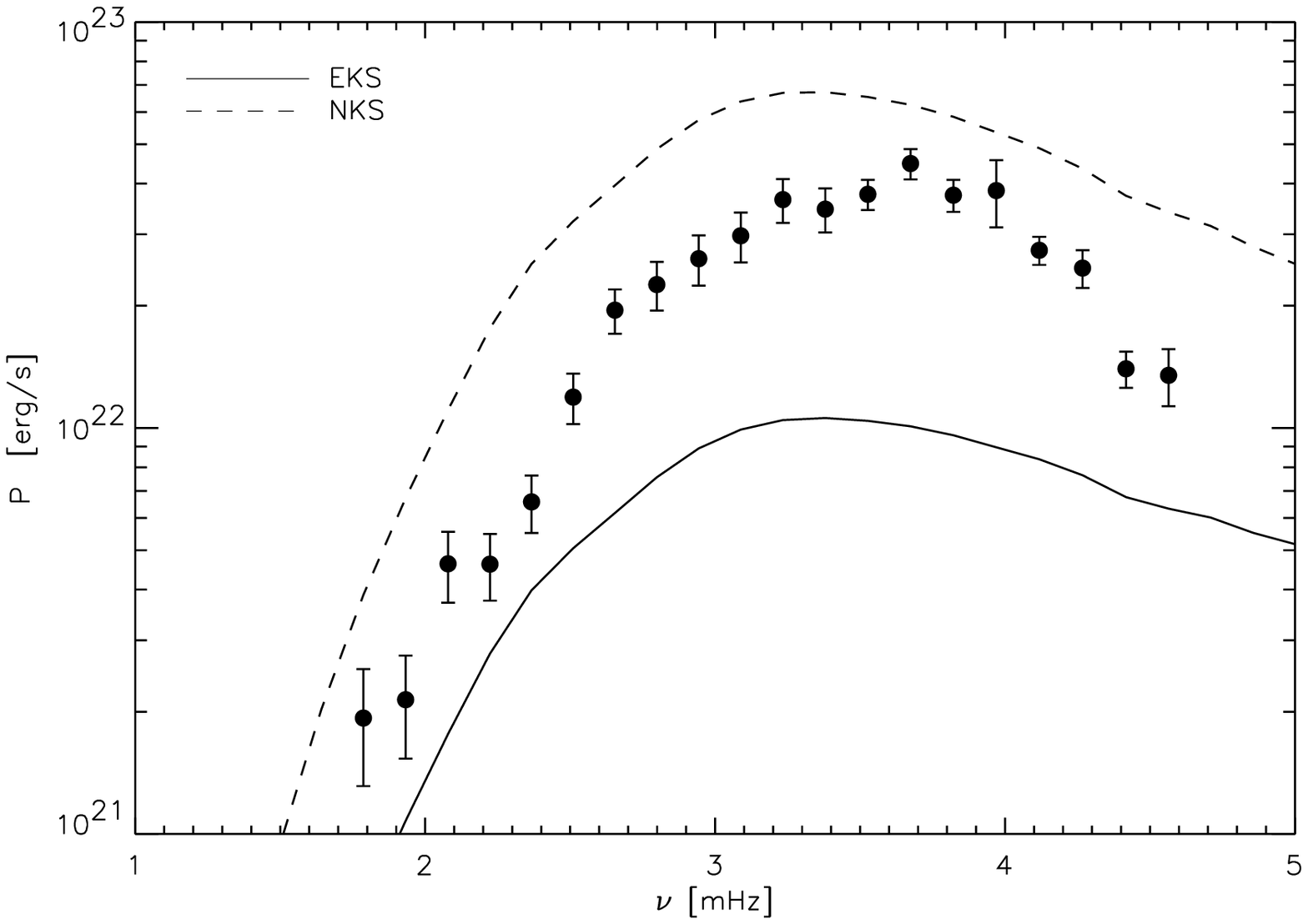}{Acoustic energy supply rate $P$ computed according to Eq.~(\ref{eqn:P}) and assuming for $E(k)$ the EKS and the NKS plotted in Fig.~\ref{spc_cinetique}. Dots represent the energy supply rate injected into the oscillations derived from the solar observations  with the help of Eq.(\ref{eqn:P_vs2}).}{Poscgh_synspc}

\subsection{Effects due to the  stratification of the turbulent spectrum at
large scale}
\label{sec:consequences:stratification of the turbulent spectrum}

\begin{figure}
\resizebox{\hsize}{!}{\includegraphics  {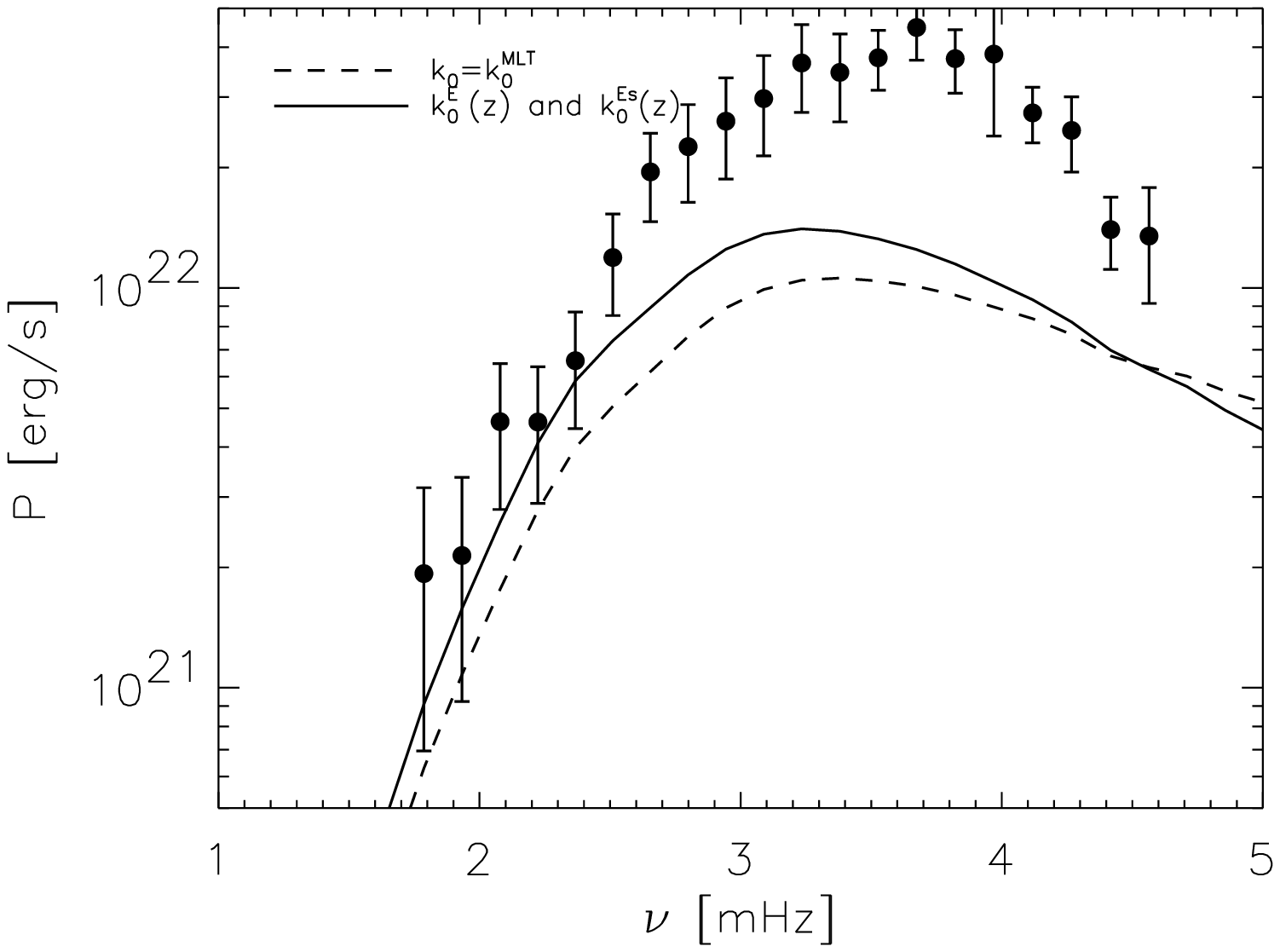}}
\caption{ 
Same as Fig.~\ref{Poscgh_MLT}.
Solid line : the  variation of $k_0^E$ and $k_0^{E_s}$ with $z$ are obtained from the simulation (see Fig.~\ref{fig:k0_z} and Sect.~\ref{sec:Turbulent spectra}).
The  dashed line  is identical to the solid line  of Fig.~\ref{Poscgh_MLT} where $k_0^E=k_0^{E_s}=k_0^{\rm MLT}(z)$.
}
\label{Poscgh_k0_k0uz}
\end{figure}

In Sect.~\ref{sec:Turbulent spectra} 
we showed  that the variations of
$k_0^{E}$ and $k_0^{E_s}$ with $z$ deduced from the 3D simulation differ 
 from the  MLT estimation as given by  Eq.~(\ref{eqn:k_0:MLT}) 
(see Fig.~\ref{fig:k0_z}). Fig.~\ref{Poscgh_k0_k0uz}
 presents the consequences 
of  the $z$  variations of $k_0^{E}$  on the oscillation amplitudes (as the variations of  $k_0^{E}$
and $k_0^{E_s}$ with depth are quite similar we assume for the sake of simplicity that $k_0^{E_s}(z)$ is equal to $k_0^{E}(z)$). 
The $z$ dependency of 
$k_0^E$ causes the maximum of $P$ to be
 larger than when assuming 
 $k_0^E=k_0^{\rm MLT}$ 
($\sim 50\,\%$ larger). 
This is due to the fact 
that  in most part of the excitation region
   $k_0^{\rm MLT}$ is smaller than $k_0^E$ and $k_0^{E_s}$ 
except above the top of the superadiabatic 
region (see Fig.~\ref{fig:k0_z}).
A larger  $k_0^E$ results in 
a larger linewidth  
$\omega_k \equiv (k u_k) / \lambda$ for $\chi_k(\omega)$ 
hence in a larger amount of acoustic energy 
injected to the mode (see Eq.~\ref{eqn:delta:omega}).

Furthermore, at high frequency, $P$ decreases with $\nu$ more rapidly than when assuming  $k_0^E=k_0^{E_s}=k_0^{\rm MLT}$.  Taking into account the actual variation $k_0^E$ with $z$  instead of assuming $k_0^E=k_0^{\rm MLT}$ makes then the $\nu$-dependency of  $P$ at high frequency closer to that of the observed excitation spectrum. This is because, above  the top of the superadiabatic  region, $k_0^E$  decreases with $z$ whereas $k_0^{\rm MLT}$  increases with $z$ . 
Indeed, the excitation of the high frequency modes occurs predominantly in the upper most part of the top of the  superadiabatic 
region. As mentionned above the line width of $\chi_k(\omega)$  decreases with decreasing  $k_0$. Therefore the contribution of the term $\chi_k(\omega)$ to the excitation of  high frequency mode is smaller when assuming the actual variation of $k_0^E$ with $z$ than  when assuming that $k_0^E$   varies as $k_0^{\rm MLT}$.

\section{Summary and discussion}
\label{sec:Conclusion}

An analysis of a 3D simulation of the upper part 
of the solar convective zone provides
time averaged  constraints upon several physical parameters
 which enter the theoretical expression for 
the supply rate of energy,  $P$,  injected
 into the solar p~modes. These  constraints  are:

1) the depth dependency: of $u^2$, the  mean square velocity- of
$w^2$, the mean square vertical component of the velocity- of  
$\tilde s^2$, the mean square values of  entropy fluctuations.

2)  the wavenumber ($k$) dependency of $E$ and $E_s$ 
the turbulent kinetic energy spectrum and  the turbulent entropy spectrum 
respectively.

3) the  depth dependency of the 
wavenumbers $k_0^E$ and $k_0^{E_s}$, 
the wavenumbers at which convective energy is maximum and 
is injected into the  turbulent inertial ranges of 
the turbulent kinetic energy spectra $E$,  $E_s$ respectively.

4) the depth dependency of $\Phi=u^2 /w^2$, the mean values of the anisotropy.

\vskip 0.5 truecm

Differences between  $w^2$ -~and $\tilde s^2$~- 
and their respective GMLT estimations  have only small consequences on 
 the profile  of the excitation rate $P(\omega)$.
However the values reached by  $w^2$ and 
$\tilde s^2$ with the 3D simulation 
are responsible for  an increase of the
 relative contribution of the Reynolds stress 
to $P(\omega)$ by a factor $\sim 1.5$ at low frequency 
$\nu \lesssim 3$~mHz compared to the one 
obtained with the GMLT solar model. 
This is because the GMLT model 
overestimates  $\tilde s^2$ by $\sim 20~\%$ at the top of excitation
 region and underestimates $w^2$ within 
 most part of the excitation region by up to $\sim 15~\%$.

\vskip 0.3 truecm

The  energy distributions $E$ and $E_s$ over eddies with
  wavenumber $k$ obtained in the simulation
scale approximately as $k^{+1}$
in the domain $k \lesssim k_0^E$. They therefore have  
approximately the same behavior
as the  'Extended Kolmogorov Spectrum'
(EKS) defined in \citet{Musielak94}.
In contrast,  their $k$-dependencies
significantly differ from  those assumed
 in the Nesis Kolmogorov Spectrum (NKS) which   
 scales as  $k^{-5}$  below $k_0^E$.
The NKS predicts much larger maximum values for $P$ 
than does the EKS.
This is because the NKS
 concentrates kinetic energy in the vicinity of $k_0^E$.

\vskip 0.3 truecm

The 3D simulation indicates  that $k_0^{E}\simeq 3.6~ {\rm Mm}^{-1}$ 
at the top of excitation region. 
This corresponds to the horizontal size of the granulation
($\sim 2$~Mm). It is worth noting that taking the depth dependency of
$k_0(z)$ into account results in a increase
of the maximum of $P$ by as much as $\sim 50 \%$ 
and brings these values even closer to the observations.
On the other hand,  only minor differences are seen 
on the frequency dependence 
of the  excitation rates $P$ when 
using the depth dependency
of $k_0(z)$ from the simulation or 
assuming the form $k_0^{\rm MLT}=2 \pi/\beta \Lambda$
with $\Lambda= \alpha H_p$ provided that $\beta$ is adjusted in 
order for $k_0^{\rm MLT}$ to match the value 
reached by $k_0^{E}$  at the layer where $w$ reaches its maximum.

\vskip 0.3 truecm

The excitation rate $P(\omega)$ is 
very sensitive to the value of $\Phi$.
In the GMLT formulation,
  the quantity $\Phi$ is a  parameter which  
is adjusted in order to obtain the best fit
 between computed solar damping rates and the solar
 measurements: the adopted value 
is  $\Phi=1.37$   \citep[see ][]{Houdek01}.
On the other hand, the 3D simulation suggests a higher value 
within the excitation region ($\Phi \simeq2$).
Larger values of  $\Phi$ result in an increase of the 
mode driving by the turbulent motions.
We find that using the value $\Phi=1.37$  underestimates 
  $P(\omega)$ by a factor 
$\sim 5$ relatively to $P(\omega)$ 
computed with $\Phi(z)$ in the excitation region from the 3D simulation.
On the other hand,  using the GMLT formulation for the convective velocity 
with  a value $\Phi\sim2 $, as suggested by the 3D simulation, yields 
a power $P(\omega)$ close to
 the one obtained by the 3D simulation.
To fix ideas,  the maximum amplitude
is $\sim 4$~cm/s, $\sim 8$~cm/s, $\sim 10$~cm/s  
when calculated with GMLT and $\Phi=1.37$, with  GMLT and $\Phi=2$
and  when using velocities and $\Phi(z)$ derived from
the 3 D simulation respectively. These figures must be compared to 
the observed maximum amplitude $\sim 23$~cm/s.

This shows  that the  values of
$\Phi$ found for the solar GMLT model when adjusted to the damping rates
 is  not compatible 
with the actual properties of the turbulent medium 
in the excitation region. 
An improvement could come from a consistent calculation 
which would  assume a depth dependent $\Phi(z)$, 
as suggested by the simulations, in both  damping rate and
excitation rate computations. 
Damping rates are indeed expected to be sensitive to depths deeper than the
excitation rate  where  smaller values of $\Phi$ are encountered
and the simulation shows that the velocity anisotropy  factor
$\Phi$ decreases from  $2$ down to 1.3 from top 
of the superadiabatic region to bottom of the simulated
solar  region.

\vskip 0.5 truecm
Without any  adjustment of scaling parameters 
but using all the constraints inferred from the 3D simulation considered here, 
we find a maximum of $P$ much 
larger ($\sim 5$ times larger) 
than the $P$ maximum obtained 
using a 1D GMLT solar model  when $\Phi$ is fixed by the observed 
damping rates.
It is also found that the so-called 
entropy source term, which arises from  
the advection of the turbulent fluctuations 
of entropy by the turbulent motions, 
is still the dominant source of the excitation.
However its contribution  to the excitation 
is now  $\sim 65-75 \%$ 
instead of $\sim 95~\%$ as found in Paper~II.

Our computation still underestimates by a factor
 $\sim 3$ the maximum value of $P$ compared with the one derived 
from the solar seismic observations by \citet{Chaplin98}.
Moreover the decrease of $P$ with $\nu$ at
 high frequency ($\nu \gtrsim 3.5$~mHz) is 
found to be significantly smaller
than the one inferred from 
the solar seismic observations, 
indicating a deficiency in the
present modelling at high frequency.

\vskip 0.5 truecm
As a final point, we discuss the  model for the 
turbulent kinetic  energy spectrum:

In Paper~II  the parameter $\lambda$ and $k_0^{E}$ were adjusted 
-~given a turbulent spectrum $E(k)$~-  
so as to obtain the best possible agreement 
between computed and measured values of the
 maximum solar oscillation amplitude
and its frequency position, as well as  the frequency-dependence of the
oscillation  amplitudes. Adjustments of these scaling  parameters led to 
a better agreement  between computed values of  $P$ and the 
seismic observations when using the NKS than the EKS.
However, the present results 
from a 3D simulation strongly suggest that the 
EKS is a better model  for the solar  turbulent kinetic  energy spectrum.

The better  agreement obtained with NKS than EKS when adjusting the free
parameters  is due to the fact that the NKS concentrates
 most of the kinetic energy in the vicinity of  $k_0^E$.
Indeed, the  NKS predominantly excites  
the modes whose period are close to the
 characteristic lifetime of the eddies of wavenumber 
$k_0^E$, i.e. the modes with frequency close to the
 frequency at which $P$ peaks ($\nu \sim 3$~mHz).
As a consequence,  the amount of energy going into the  
high frequency modes is  relatively smaller with the NKS 
than it is with the EKS. This  explains why the NKS reproduces 
better the steep decrease with  $\nu$ of $P$ 
at high frequency and results in a value
 for  $k_0^E$ identical to that inferred 
from the simulation ($k_0^{E}\simeq 3.6~ {\rm Mm}^{-1}$). 
In contrast, whatever the adjustment, 
the EKS  reproduces  neither 
the  frequency dependence of $P$ at high frequency nor
the value $k_0^{E}\simeq 3.6~ {\rm Mm}^{-1}$.  
Hence assuming that the 3D simulation yields 
the proper behavior of the solar 
 kinetic energy spectrum, well modelled by the EKS, 
one is led to conclude that the excitation as given by the present 
stochastic excitation model is not  efficient enough  at
large scales ($k \sim k_0$) and too efficient at small scales ($k > k_0$). 

\vskip 0.5 truecm

Discrepancies between our calculations  and the observed excication rates or the results by \citet{Stein01II} are 
likely due  to dynamic properties of 
turbulence which are not properly taken into account in the excitation model.
Indeed, the dynamic properties of turbulence are modeled by the function $ \chi_k$. All current theoretical calculations of the excitation rates assume a gaussian function for $ \chi_k$  (e.g. \citet[GK hereafter]{GK77}, \citet{Balmforth92c}).
The gaussian model is likely to be at the origin of the current under-estimate of the rates at which solar $p$-modes are excited \citep[see forthcoming paper][]{Samadi02II}.
This  may also explain the fact that we find that the 
entropy source term is dominant over the Reynolds stress 
contribution whereas \citet{Stein01II} in their 
direct computations found the reverse.  
In a recent study based on a frequency analysis of the present simulation we investigate what model can correctly reproduce  model  $\chi_k$ in the frequency range where the acoustic energy injected into the solar $p$-modes is important   \citep[see forthcoming paper][]{Samadi02II}.

\vskip 0.5 truecm

In the manner of \citet{Rosenthal99} constraints from 3D simulation can be imposed to the 1D model.
According to the authors, such constraints result in a better agreement between the observed frequencies of the solar $p$-modes and the eigenfrequencies of the computed adiabatic oscillations.
An improvement  in the  calculation of the excitation rates at solar-type oscillations could then also  come from a more consistent calculation of the eigenmodes  which would use such constrained 1D model. 

\begin{acknowledgements}
We thank H.-G. Ludwig for valuable help in analyzing the simulated data. 
We are indebted to G. Houdek for providing us the solar model. 
We thank the  referee (B. Dintrans) for his  meaningful comments.
RS's work has been supported in part by the Particle Physics and 
Astronomy Research Council of the UK under grant PPA/G/O/1998/00576.
\end{acknowledgements}



\begin{thebibliography}{32}
\expandafter\ifx\csname natexlab\endcsname\relax\def\natexlab#1{#1}\fi

\bibitem[{{Baglin} \& {The Corot Team}(1998)}]{Baglin98}
{Baglin}, A. \& {The Corot Team}. 1998, in IAU Symp. 185: New Eyes to See
  Inside the Sun and Stars, Vol. 185, 301

\bibitem[{{Balmforth}(1992{\natexlab{a}})}]{Balmforth92c}
{Balmforth}, N.~J. 1992{\natexlab{a}}, \mnras, 255, 639

\bibitem[{{Balmforth}(1992{\natexlab{b}})}]{Balmforth92a}
---. 1992{\natexlab{b}}, \mnras, 255, 603

\bibitem[{{B\"ohm - Vitense}(1958)}]{Bohm58}
{B\"ohm - Vitense}, E. 1958, Zeitschr. Astrophys., 46, 108

\bibitem[{{Canuto} {et~al.}(1996){Canuto}, {Goldman}, \&
  {Mazzitelli}}]{Canuto96}
{Canuto}, V.~M., {Goldman}, I., \& {Mazzitelli}, I. 1996, \apj, 473, 550

\bibitem[{{Canuto} \& {Mazzitelli}(1991)}]{Canuto91}
{Canuto}, V.~M. \& {Mazzitelli}, I. 1991, \apj, 370, 295

\bibitem[{{Chaplin} {et~al.}(1998){Chaplin}, {Elsworth}, {Isaak}, {Lines},
  {McLeod}, {Miller}, \& {New}}]{Chaplin98}
{Chaplin}, W.~J., {Elsworth}, Y., {Isaak}, G.~R., {et~al.} 1998, \mnras, 298,
  L7

\bibitem[{{Christensen-Dalsgaard}(1982)}]{JCD82}
{Christensen-Dalsgaard}, J. 1982, \mnras, 199, 735

\bibitem[{{Eggleton} {et~al.}(1973){Eggleton}, {Faulkner}, \&
  {Flannery}}]{Eggleton73}
{Eggleton}, P.~P., {Faulkner}, J., \& {Flannery}, B.~P. 1973, \aap, 23, 325

\bibitem[{{Favata} {et~al.}(2000){Favata}, {Roxburgh}, \&
  {Christensen-Dalsgaard}}]{Favata00}
{Favata}, F., {Roxburgh}, I., \& {Christensen-Dalsgaard}, J. 2000, in The Third
  MONS Workshop : Science Preparation and Target Selection, 49--54

\bibitem[{{Goldreich} \& {Keeley}(1977)}]{GK77}
{Goldreich}, P. \& {Keeley}, D.~A. 1977, \apj, 212, 243

\bibitem[{{Goldreich} {et~al.}(1994){Goldreich}, {Murray}, \& {Kumar}}]{GMK94}
{Goldreich}, P., {Murray}, N., \& {Kumar}, P. 1994, \apj, 424, 466

\bibitem[{{Gough}(1977)}]{Gough77}
{Gough}, D.~O. 1977, \apj, 214, 196

\bibitem[{{Houdek}(1996)}]{Houdek96}
{Houdek}, G. 1996, PhD thesis, Institut f\"ur Astronomie , Wien

\bibitem[{{Houdek} {et~al.}(1999){Houdek}, {Balmforth},
  {Christensen-Dalsgaard}, \& {Gough}}]{Houdek99}
{Houdek}, G., {Balmforth}, N.~J., {Christensen-Dalsgaard}, J., \& {Gough},
  D.~O. 1999, \aap, 351, 582

\bibitem[{{Houdek} {et~al.}(2001){Houdek}, {Chaplin}, {Appourchaux},
  {Christensen-Dalsgaard}, {D{\" a}ppen}, {Elsworth}, {Gough}, {Isaak}, {New},
  \& {Rabello-Soares}}]{Houdek01}
{Houdek}, G., {Chaplin}, W.~J., {Appourchaux}, T., {et~al.} 2001, \mnras, 327,
  483

\bibitem[{{Houdek} \& {Gough}(2002)}]{Houdek02}
{Houdek}, G. \& {Gough}, D.~O. 2002, \mnras, 336, L65

\bibitem[{{Musielak} {et~al.}(1994){Musielak}, {Rosner}, {Stein}, \&
  {Ulmschneider}}]{Musielak94}
{Musielak}, Z.~E., {Rosner}, R., {Stein}, R.~F., \& {Ulmschneider}, P. 1994,
  \apj, 423, 474

\bibitem[{{Nesis} {et~al.}(1993){Nesis}, {Hanslmeier}, {Hammer}, {Komm},
  {Mattig}, \& {Staiger}}]{Nesis93}
{Nesis}, A., {Hanslmeier}, A., {Hammer}, R., {et~al.} 1993, \aap, 279, 599

\bibitem[{{Nordlund} {et~al.}(1997){Nordlund}, {Spruit}, {Ludwig}, \&
  {Trampedach}}]{Nordlund97}
{Nordlund}, A., {Spruit}, H.~C., {Ludwig}, H.~G., \& {Trampedach}, R. 1997,
  \aap, 328, 229

\bibitem[{{Osaki}(1990)}]{Osaki90}
{Osaki}, Y. 1990, in Lecture Notes in Physics : Progress of Seismology of the
  Sun and Stars, ed. Y.~{Osaki} \& H.~{Shibahashi} (Springer-Verlag), 75

\bibitem[{{Rieutord} {et~al.}(2000){Rieutord}, {Roudier}, {Malherbe}, \&
  {Rincon}}]{Rieutord00}
{Rieutord}, M., {Roudier}, T., {Malherbe}, J.~M., \& {Rincon}, F. 2000, \aap,
  357, 1063

\bibitem[{{Rosenthal} {et~al.}(1999){Rosenthal}, {Christensen-Dalsgaard},
  {Nordlund}, {Stein}, \& {Trampedach}}]{Rosenthal99}
{Rosenthal}, C.~S., {Christensen-Dalsgaard}, J., {Nordlund}, {\AA}., {Stein},
  R.~F., \& {Trampedach}, R. 1999, \aap, 351, 689

\bibitem[{{Samadi}(2001)}]{Samadi01}
{Samadi}, R. 2001, in SF2A-2001: Semaine de l'Astrophysique Francaise, E148
  (astro--ph/0108363)

\bibitem[{{Samadi} {et~al.}(2003){Samadi}, { Nordlund}, {Stein}, {Goupil}, \&
  {Roxburgh}}]{Samadi02II}
{Samadi}, R., { Nordlund}, {\AA}., {Stein}, R., {Goupil}, M.-J., \& {Roxburgh},
  I. 2003, submitted to \aap

\bibitem[{{Samadi} \& {Goupil}(2001)}]{Samadi00I}
{Samadi}, R. \& {Goupil}, M.~. 2001, \aap, 370, 136 (Paper~I)

\bibitem[{{Samadi} {et~al.}(2001){Samadi}, {Goupil}, \&
  {Lebreton}}]{Samadi00II}
{Samadi}, R., {Goupil}, M.~., \& {Lebreton}, Y. 2001, \aap, 370, 147 (Paper~II)

\bibitem[{{Samadi} {et~al.}(2002){Samadi}, {Houdek}, {Goupil}, \&
  {Lebreton}}]{Samadi00III}
{Samadi}, R., {Houdek}, G., {Goupil}, M.-J., \& {Lebreton}, Y. 2002, submitted
  to \aap (Paper~III)

\bibitem[{{Stein}(1967)}]{Stein67}
{Stein}, R.~F. 1967, Solar Physics, 2, 385

\bibitem[{{Stein} \& {Nordlund}(1998)}]{Stein98}
{Stein}, R.~F. \& {Nordlund}, A. 1998, \apj, 499, 914

\bibitem[{{Stein} \& {Nordlund}(2001)}]{Stein01II}
{Stein}, R.~F. \& {Nordlund}, {\AA}. 2001, \apj, 546, 585

\bibitem[{{Unno} \& {Spiegel}(1966)}]{Unno66}
{Unno}, W. \& {Spiegel}, E.~A. 1966, \pasj, 18, 85

\end{thebibliography}
\bibliographystyle{aa}

\end{document}